\documentclass[aps,prd]{revtex4}
\usepackage{amssymb,amsmath,graphicx,epsfig,bm,color}
\usepackage{leftidx}
\usepackage{booktabs}
\usepackage{multirow}
\usepackage{amsmath,amssymb}
\usepackage{mathtools}
\usepackage{float}
\usepackage{color}
\usepackage{leftidx}
\usepackage{booktabs}
\usepackage{latexsym}
\usepackage{revsymb}
\usepackage{multirow}
\usepackage{array}
\newcommand{\be}{\begin{eqnarray}}
\newcommand{\ee}{\end{eqnarray}}
\usepackage[colorlinks]{hyperref}
\hypersetup{
    colorlinks=red,
    citecolor=blue,
    filecolor=magenta,      
    urlcolor=red,
}
\renewcommand{\d}{\mathrm{d}}
\newcommand{\pup}{p^\uparrow}

\begin{document}

\title{Sivers Asymmetry in  Photoproduction of $J/\psi$ and Jet  at the EIC }

\author{ Raj Kishore}

\affiliation{ Department of Physics,
Indian Institute of Technology Bombay,Powai, Mumbai 400076,
India}

\author{ Asmita Mukherjee}
\affiliation{ Department of Physics,
Indian Institute of Technology Bombay,Powai, Mumbai 400076,
India}
\author{  Sangem Rajesh}

\affiliation{Dipartimento di Fisica, Universit\`a di Cagliari, Cittadella Universitaria, I-09042 Monserrato (CA), Italy}
\affiliation{INFN, Sezione di Cagliari, Cittadella Universitaria, I-09042 Monserrato (CA), Italy}

\date{\today}

\date{\today}
\begin{abstract}
We calculate the Sivers asymmetry in the photoproduction of almost back-to-back  $J/\psi$-jet pair 
in the process $ep^\uparrow \to J/\psi+\mathrm{jet}+X$, which will be
possible   at the future planned electron-ion collider (EIC). We use the framework of generalized 
parton model
(GPM), and NRQCD for calculating the $J/\psi$ production rate. We include
contributions from both color singlet and color octet states in the
asymmetry.  We obtain sizable Sivers asymmetry that can be promising to
determine the gluon Sivers function. We also investigate the effect of TMD
evolution on the asymmetry.  
\end{abstract}
\maketitle
%\raggedbottom 
\section{Introduction}\label{sec1}

Single spin asymmetries and transverse momentum dependent parton distributions (TMD pdfs) are objects of a lot of interest in 
recent days in hadron physics. Among the TMD pdfs Sivers function \cite{Sivers:1989cc} is of particular interest. This gives 
the distribution of unpolarized quarks/gluons in a transversely polarized nucleon, which is not left-right symmetric with respect
to the plane formed by the transverse momentum and spin of the nucleon. In some model calculations Sivers function is shown to be
related to the quark orbital angular momentum through the GPD $E_q$ \cite{Burkardt:2003uw}.  Sivers function introduces an asymmetry, 
for example,  in the azimuthal angle of the observed final state hadron in semi-inclusive deep inelastic scattering (SIDIS) and in
the azimuthal angle correlations of the lepton pair in Drell-Yan process or back-to-back jets in $pp$ collision, called the Sivers asymmetry. 
The first transverse moment of the Sivers function is related to the twist-three Qiu-Sterman function \cite{Qiu:1991pp}. 
First experimental information on non-zero Sivers function for quarks was obtained from HERMES \cite{Airapetian:2004tw} 
and COMPASS \cite{Adolph:2012sp} results. Since then, quite a lot of advances have been made, both in theory and experiment.
Parametrization of quark Sivers function has been obtained in \cite{Anselmino:2016uie} and for gluons in
\cite{DAlesio:2015fwo, DAlesio:2018rnv}
by fitting data from RHIC within the DGLAP evolution approach. TMDs evolve with scale in 
a different way compared to the collinear pdfs. Much progress have been made in 
the past few years to understand the TMD evolution \cite{Aybat:2011ge, Aybat:2011zv, Collins:2017oxh},
and unpolarized distributions 
and fragmentation function have been calculated at NNLO  \cite{Echevarria:2016scs}. 
A parametrization of the Sivers function
incorporating the TMD evolution has been obtained in \cite{Echevarria:2014xaa}, 
however, gluon Sivers function (GSF) is not yet known. Therefore, compared to the quark TMDs, gluon TMDs
are much less known, and these will also be investigated at the future electron-ion collider (EIC) \cite{Accardi:2012qut}
and the future fixed target plans at the LHC \cite{Brodsky:2012vg, Kikola:2017hnp, Trzeciak:2017csa}. Gluon TMDs satisfy the 
positivity bounds first derived in \cite{Mulders:2000sh}. A phenomenological bound on
GSF 
was obtained \cite{Burkardt:2004ur}, commonly known as Burkardt sum rule, from the requirement that the net transverse momentum 
of all  partons (quarks and gluons) in a transversely polarized nucleon should vanish. In \cite{Anselmino:2008sga}, a fit to the 
data from SIDIS at low scale indicates that this sum rule is almost saturated by contribution from u and d quarks, however, 
there may still be about $30 \%$ contribution from GSF. \par
Sivers function is a T-odd object and  initial and final state interactions play an important role in Sivers asymmetry \cite{Brodsky:2002rv}.
They are resummed into the gauge link or Wilson line in the operator definition of the Sivers function that is needed for color gauge invariance
\cite{Boer:2003cm}. Gluon TMDs have two gauge links, in contrast to quark TMDs, that have only one. This introduces process dependence in them.
The Sivers function in SIDIS is expected to be equal in magnitude but opposite in sign compared to the Sivers function appearing in the
Drell-Yan process \cite{Brodsky:2002rv}.  Recent  data from RHIC \cite{Adamczyk:2015gyk} as well as COMPASS \cite{Aghasyan:2017jop} seem
to favour the sign change, however more data are needed \cite{Anselmino:2016uie} . The GSF for any process, in general 
can be written in terms of two independent functions,  one of them has an operator structure that is C-even, and the other, C-odd. 
In the literature, the former is called a f-type Sivers function and the latter, d-type \cite{Bomhof:2006ra, Buffing:2013kca}. 
In fact, one of these two (d-type) is not constrained by the Burkardt sum rule. More experimental data are needed to precisely 
determine the GSF.  As is well known, $J/\psi$ production in $ep$  and $pp$ collision is an effective method to
probe the gluon TMDs including the GSF \cite{ Kishore:2018ugo, Lansberg:2017tlc, Rajesh:2018qks, Mukherjee:2016qxa, DAlesio:2017rzj, Mukherjee:2015smo, Mukherjee:2016cjw}, as contribution 
to the Sivers asymmetry comes already at leading order (LO) through the virtual photon-gluon or gluon-gluon fusion processes, respectively.
Data on Sivers asymmetry in  $J/\psi$ production  are available from COMPASS collaboration \cite{Matousek:2016xbl}, although with large error
bars, it can be qualitatively explained by a LO calculation in NRQCD based color octet model \cite{Mukherjee:2016qxa}. 
In a recent work \cite{DAlesio:2019qpk}, maximal values of  the azimuthal asymmetries in back-to-back electroproduction of $J/\psi$ and
a jet is estimated within the TMD factorization framework by neglecting the intrinsic transverse momentum of the initial parton in the hard part.
Another interesting process to probe the GSF is quasi-real photoproduction of  a hadron \cite{DAlesio:2017nrd} or $J/\psi$ \cite{Godbole:2012bx,Godbole:2013bca,Rajesh:2018qks}. 
Contribution to the single spin asymmetry  (SSA) comes from $J/\psi$ observed in the forward region that is when the transverse momentum ($p_T$)
of $J/\psi$ is small.   In this work, we investigate the possibility to probe the GSF in quasi-real photoproduction 
of back-to-back  $J/\psi$ and jet by employing the generalized parton model (GPM) wherein the intrinsic transverse momentum of the initial parton is 
considered in the hard part, which will be possible in the future EIC. This will be sensitive to the GSF in a different kinematical region,
and $J/\psi$ observed need not be in the forward region. \par
As mentioned  above, initial and final state interactions play a very important role in  the SSAs.  TMD factorization has been proven 
only for certain processes. The current status of the TMD factorization for heavy quarkonium  production in  $pp$ collisions can be found 
in \cite{Echevarria:2019ynx}.  The most widely used approach to calculate amplitudes for $J/\psi $
production is based on non-relativistic QCD (NRQCD). In this approach, the amplitude is factorized into a soft non-perturbative part
and a hard part \cite{Carlson:1976cd,Berger:1980ni,Baier:1981uk,Baier:1981zz,Braaten:1994vv,Cho:1995vh,Cho:1995ce}. The heavy quark pair 
is produced in color singlet (CS) or color octet (CO) states in hard interaction. This is calculated in the perturbation theory. Then this heavy
quark pair hadronizes to a quarkonium by emitting soft gluons. The hadronization process is described in terms of long distance matrix elements
(LDMEs), that are obtained by fitting experimental data. The LDMEs have definite scaling properties with respect to the velocity parameter 
$v$, which is assumed to be small $v << 1$ \cite{Lepage:1992tx}. The theoretical estimates are  arranged in  a double expansion in powers 
of $v$ and the strong coupling, $\alpha_s$. NRQCD has been successful in explaining hadroproduction data from
TEVATRON \cite{Abe:1997jz,Acosta:2004yw} and also $J/\psi$ photoproduction data from HERA
\cite{Adloff:2002ex,Aaron:2010gz,Chekanov:2002at,Abramowicz:2012dh}. Both CS and CO contributions are needed to explain the HERA data 
\cite{Rajesh:2018qks}. In this work we calculate the weighted Sivers asymmetry, $A_N^{\sin(\phi_q)}$, in photoproduction of back-to-back $J/\psi$ and jet at EIC in
NRQCD by incorporating both CS and CO states. The plan of the paper is as follows.
In section \ref{sec2} we give the analytic expressions of  the asymmetry. In section \ref{sec3} we present the TMD evolution framework. 
Numerical results and  conclusion are given in section \ref{sec4} and  \ref{sec5} respectively.    

%----------------------------------------------------------------------------------------
\section{Sivers Asymmetry }\label{sec2}

\begin{figure}[h]
\begin{center} 
\includegraphics[height=7cm,width=11cm]{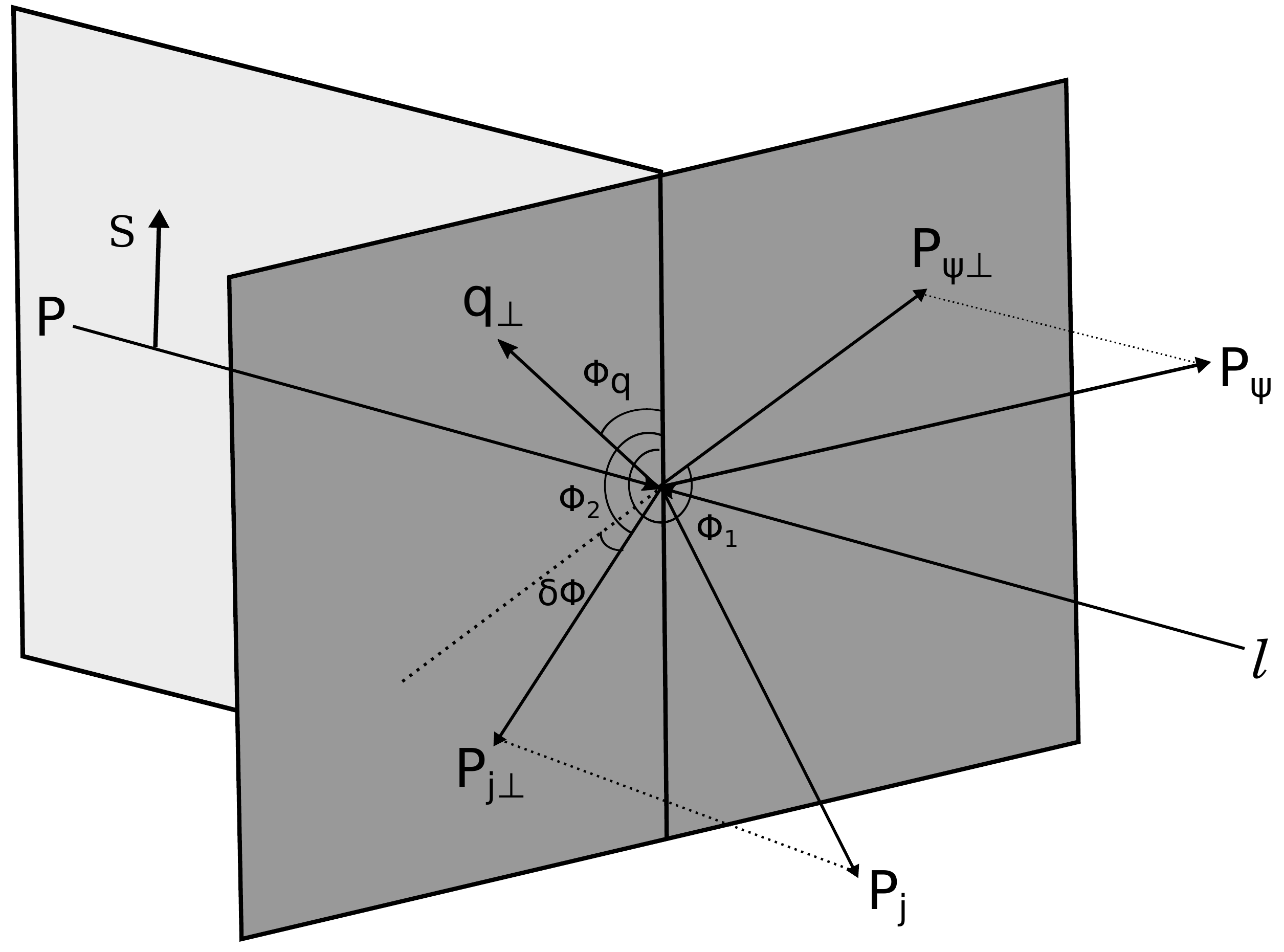}
\end{center}
\caption{\label{fig1} Illustration of azimuthal angles in the $ep^\uparrow\rightarrow J/\psi+$jet process.
The transverse momenta  $\bm P_{\Psi \perp}$ and $\bm P_{j \perp}$ of $J/\psi$ and jet respectively
are in the plane orthogonal to the momentum of the proton $P$.}
\end{figure}

We consider the photoproduction process 
\begin{equation}
 e(l)+p^\uparrow(P) \rightarrow J/\psi (P_\psi)+\mathrm{jet}(P_{j})+X,
\end{equation}
where the arrow in the superscript indicates the polarization of the proton. The letters in the round 
brackets represent the four momentum of the corresponding particles. We consider the proton-electron
center of mass (C.M) frame wherein the proton and electron move along the +$z$ and -$z$ direction.
 The transverse plane, defined in \figurename{~\ref{fig1}}, is orthogonal to the momentum of proton direction.
The initial scattering electron radiates the virtual photon that will interact with the proton.
The four momentum square of the virtual photon is  $q^2\approx -2E_eE_e^\prime(1-\cos\theta)$ with 
$E_e$ and $E_e^\prime$ are energies of the initial and final scattered electron respectively.
In the forward scattering limit, photoproduction, the four momentum of the virtual photon $q^2=-Q^2\rightarrow 0$ as a 
result the virtual photon becomes the real  photon.
The dominant subprocess for $J/\psi$
production is the $\gamma (q)+ g(p)\rightarrow J/\psi(P_\psi)+g (P_{j}) $ at next-to-leading order (NLO) in $\alpha_s$.
The quark (antiquark) initiated subprocess can also contribute $\gamma +q ( \bar q) 
\rightarrow J/\psi+q (\bar q) $.
The unpolarized differential cross section for $ep \rightarrow J/\psi+\mathrm{jet}+X$ process 
can be written as follows
\begin{equation}\label{eqd1}
 \begin{aligned}
E_\Psi E_j\frac{d\sigma}{d^3{\bm P}_\Psi d^3{\bm P}_j}=\frac{d\sigma}{d^2{\bm P}_{\Psi \perp}dz
d^2{\bm P}_{j\perp}dz_1}
={}&\frac{1}{4(2\pi)^2}\frac{1}{zz_1}{\sum_{a}}\int dx_\gamma dx_a 
d^2{\bm p}_{a\perp}
f_{\gamma/e}(x_\gamma)f_{a/p}(x_a,{p}_{a\perp })\\
&\times\delta^4(q+p-P_\Psi-P_j)\frac{1}{2\hat{s}}
|\mathcal{M}_{\gamma a\rightarrow J/\psi a}|^2,
\end{aligned}
\end{equation}
{where $a=g,u,d,s,\bar{u},\bar{d},\bar{s}$}. The $x_\gamma$ and $x_a$ are the
light-cone momentum fractions of photon and partons respectively, 
and ${\bm p}_{a\perp }$ is the transverse momentum of the initial parton.  We have assumed TMD 
factorization in the GPM model 
with the inclusion of intrinsic transverse momentum of the initial parton in the hard part. When the exchanged photon is quasi-real, the inreaction takes place through the 
Weizs$\ddot{a}$ker-Williams distribution function of the electron, $f_{\gamma/e}(x_\gamma)$, this describes the density of photons inside the electron
and is given by \cite{Frixione:1993yw}
\begin{eqnarray}  \label{flux}
f_{\gamma/e}(x_{\gamma})=\frac{\alpha}{2\pi }\left[
2m_e^2x_\gamma \left(\frac{1}{Q^2_{min}}-\frac{1}{Q^2_{max}}\right)+\frac{1+(1-x_\gamma)^2}{x_\gamma}
\ln\frac{Q^2_{max}}{Q^2_{min}}\right ],
\end{eqnarray}
where $\alpha$ is the electromagnetic coupling and $Q^2_{min}=m_e^2\frac{x_\gamma^2}{1-x_\gamma}$, $m_e$ 
being the electron mass. We have considered $Q^2_{max}=1$ GeV$^2$ for estimating the Sivers asymmetry.
The unpolarized  TMD, $f_{a/p}$, represents the density of unpolarized partons inside an
unpolarized proton with momentum fraction $x_a$ and transverse momentum $p_{a\perp}$. The 
$\hat{s}$, $\hat{t}$ and $\hat{u}$ are the Mandelstam variables at partonic level and their definitions are given in appendix 
\ref{ap1}. $\mathcal{M}_{\gamma a\rightarrow J/\psi a}$ is the  amplitude of gluon and quark (anti-quark)
initiated subprocesses. The square of the amplitude is calculated using the NRQCD model,
for more details Ref.\cite{Rajesh:2018qks} is  referred  for gluon channel, and the quark  (anti-quark) channel matrix elements are given in the appendix \ref{ap2}.
The CS and CO states i.e.,  $\leftidx{^{3}}{S}{_1}^{(1,8)}$,  
$\leftidx{^{1}}{S}{_0}^{(8)}$ and $\leftidx{^{3}}{P}{_J}^{(8)}$ are considered  for $J/\psi$ production.
The center of mass (C.M) energy of the proton-electron 
system is $s=(P+l)^2$. In Eq.\eqref{eqd1}, the inelastic variables $z=\frac{P\cdot P_h}{P\cdot q}$ and
$z_1=\frac{P\cdot P_j}{P\cdot q}$ are the energy fractions transferred from photon to $J/\psi$ and jet respectively in the proton
rest frame.
In photoproduction, the inelastic variables $z$ and $z_1$  can be measured in experiments using the 
Jacquet-Blondel method 
\cite{Adloff:2002ex,Aaron:2010gz,Abramowicz:2012dh}.
By using the definitions of four momenta as given in appendix \ref{ap1}, the momentum conservation 
delta function can be decomposed as
\begin{equation}\label{delta}
 \delta^4(q+p-P_\Psi-P_j)=\frac{2}{x_\gamma s}\delta\left(1-z-z_1\right)\delta\left(x_a-\frac{M^2+
 P^2_{\Psi \perp}}{z x_\gamma s}-\frac{P^2_{j \perp}}{z_1 x_\gamma s}\right)\delta^2({\bm 
p}_{a \perp}-{\bm P}_{\Psi \perp}-{\bm P}_{j \perp}),
\end{equation}
where ${\bm P}_{\Psi \perp}$ and ${\bm P}_{j \perp}$ are the transverse momentum of the $J/\psi$ and jet
respectively and their azimuthal angles are represented with $\phi_1$ and $\phi_2$ such that
$\delta\phi=\phi_2-\phi_1-\pi$ as shown in \figurename{~\ref{fig1}}.\par
%=======================================================
%===========================================
Now, we define the sum and difference of transverse momenta of $J/\psi$ and jet  as 
${\bm q}_\perp={\bm P}_{\Psi \perp}+{\bm P}_{j \perp} $ and ${\bm K}_\perp=({\bm P}_{\Psi \perp}
-{\bm P}_{j \perp})/2$. We are interested in the 
case where $|{\bm q}_\perp|\ll|{\bm K}_\perp|$ i.e., the $J/\psi$ and jet are almost back to back in the 
transverse plane as shown in \figurename{~\ref{fig1}}.   
The azimuthal angle of $\bm q_{\perp}$ is denoted with $\phi_q$.
After integrating over $z_1$, $x_a$ and ${\bm p}_{a\perp}$,   one obtain the following expression
\begin{equation}\label{d1}
 \begin{aligned}
\frac{d\sigma}{d^2{\bm q}_{ \perp}dz
d^2{\bm K}_{\perp}}
={}&\frac{1}{2(2\pi)^2}\frac{1}{z(1-z)s}{\sum_{a}}\int \frac{dx_\gamma}{x_\gamma} 
f_{\gamma/e}(x_\gamma)f_{a/p}(x_a,{q}_{\perp })\frac{1}{2\hat{s}}
|\mathcal{M}_{\gamma a\rightarrow J/\psi a}|^2.
\end{aligned}
\end{equation}
For a tranversly polarized proton, the differential cross section is given by

\begin{equation}
\begin{aligned}
\frac{d\sigma^{\uparrow(\downarrow)}}{d^2{\bm q}_{ \perp}dz
	d^2{\bm K}_{\perp}}
={}&\frac{1}{2(2\pi)^2}\frac{1}{z(1-z)s}{\sum_{a}}\int \frac{dx_\gamma}{x_\gamma} 
f_{\gamma/e}(x_\gamma)f_{a/p^{\uparrow(\downarrow)}}(x_a,{q}_{\perp })\frac{1}{2\hat{s}}
|\mathcal{M}_{\gamma a\rightarrow J/\psi a}|^2.
\end{aligned}
\end{equation}

 The weighted Sivers asymmetry is defined as \cite{Boer:2016fqd} 
\begin{equation}\label{eq:AN}
A^{W(\phi_q)}_N  \equiv  \frac{\int d\phi_qW(\phi_q)
(\d \sigma^\uparrow - \d \sigma^\downarrow)}{\int d\phi_q(\d \sigma^\uparrow + \d \sigma^\downarrow)}
 \equiv \frac{\int d\phi_qW(\phi_q)\d\Delta\sigma(\phi_q)}{\int d\phi_q 2 \d\sigma},
\end{equation}
where $\d \sigma^{\uparrow(\downarrow)}$ indicates the polarized cross section in the process where one 
of the initial particle is transversely polarized with respect to its momentum direction. 
The azimuthal weight factor $W(\phi_q)= -\sin(\phi_q)$
which is given by  \cite{Bacchetta:2004jz,Bacchetta:2007sz}
\begin{eqnarray}
% \cos(\phi_q)=\frac{(\hat{\bm P}\times \bm S)}{|\hat{\bm P}\times {\bm S}|}.
% \frac{(\hat{\bm P}\times {\bm q}_\perp)} {|\hat{\bm P}\times {\bm q}_\perp|}\nonumber \\
 -\sin(\phi_q)=\frac{(\bm S\times \hat{\bm P} )\cdot{\bm q}_\perp}{|{\bm S}\times\hat{\bm P} |
 |\hat{\bm P}\times {\bm q}_\perp|}.
\end{eqnarray}
It has been advertised that the $J/\psi$ production probes the gluon TMDs and the quark contribution can be
safely neglected in the kinematical region considered 
because of the insignificant contribution of quarks w.r.t gluon, which will be discussed
in the results section. Hence the 
dominant contribution to the Sivers asymmetry comes from the gluon channel.
The numerator of the asymmetry is sensitive to the  Sivers function  in $J/\psi$ production

 \begin{align}
 \d\Delta\sigma \equiv & \frac{d\sigma^\uparrow}{d^2{\bm q}_{ \perp}dz
d^2{\bm K}_{\perp}} -
\frac{d\sigma^\downarrow}{d^2{\bm q}_{ \perp}dz
d^2{\bm K}_{\perp}} =\frac{1}{2(2\pi)^2}\frac{1}{z(1-z)s}{\sum_a}\int \frac{dx_\gamma}{x_\gamma}
f_{\gamma/e}(x_\gamma)\Delta\hat f_{a/p}(x_a,{\bm q}_{\perp })\frac{1}{2\hat{s}}
|\mathcal{M}_{\gamma a\rightarrow J/\psi a}|^2,
\end{align}
with $\Delta\hat f_{a/{p^\uparrow}}(x_a, {\bm q}_{\perp })$ being the  Sivers function,
describes the number density
of unpolarized partons in a transversely polarized proton with mass $M_p$. The analytic expressions for contributions from different states can be found in \cite{Rajesh:2018qks}. As only $J/\psi$ was observed there, we had integrated over the phase space of the final gluon $a=g$, whereas here, the final parton is producing the observed jet.  Sivers function in Trento 
convention \cite{Bacchetta:2004jz} is given by
\begin{align}
\Delta \hat f_{a/\pup}\,(x_a, \bm q_{\perp })  \,&\equiv 
\hat f_{a/\pup}\,(x_a, \bm q_{\perp}) - \hat f_{a/p^\downarrow}\,
(x_a, \bm q_{\perp })\nonumber \\
\label{defsiv}
&= \Delta^N f_{a/\pup}\,(x_a, q_{\perp }) \hat{\bm S}\cdot(\hat{P}\times\hat{\bm q}_\perp)\nonumber \\
&= -\Delta^N f_{a/\pup}\,(x_a, q_{\perp }) \sin(\phi_q)\nonumber \\
&= -\frac{2}{M_p} \, f_{1T}^{\perp } (x_a, q_{\perp }) \hat{\bm S}\cdot(\hat{P}\times{\bm q}_\perp).
\end{align}
The Sivers function fulfills the following positivity bound
\begin{equation}
\vert \Delta^N f_{a/\pup}\,(x_a, q_{\perp }) \vert  \le 2\,f_{a/p}\,(x_a, q_{\perp })
\,,~~{\mathrm{ or}}~~
\frac{q_\perp}{M_p}\, \vert f_{1T}^{\perp } (x_a, q_{\perp })\vert \le  f_{a/p}\,(x_a, q_{\perp })~.
\end{equation}
Following Ref.\cite{DAlesio:2015fwo,DAlesio:2018rnv}, we adopt the Gaussian
parametrization for Sivers function within the DGLAP evolution approach as  given below
\begin{equation}\label{eq:siv-par-1}
\Delta^N\! f_{a/p^\uparrow}(x_a,q_{\perp }) =   \left (-2\frac{q_\perp}{M_p}  \right )f_{1T}^{\perp} 
(x_a,q_{\perp })  = 2 \, {\cal N}_a(x_a)\,f_{a/p}(x_a)\,
h(q_{\perp })\,\frac{e^{-q_{\perp }^2/\langle q_{\perp }^2 \rangle}}
{\pi \langle q_{\perp }^2 \rangle}\,,
\end{equation}
where $f_{a/p}(x_a)$ is the usual  collinear parton distribution function (PDF) which follows the  DGLAP evolution equation and
\begin{equation}\label{ngx}
{\cal N}_a(x_a) = N_a x_a^{\alpha}(1-x_a)^{\beta}\,
\frac{(\alpha+\beta)^{(\alpha+\beta)}}
{\alpha^{\alpha}\beta^{\beta}},
\end{equation}
with $|N_a|\leq 1$ and
\begin{equation}
h(q_\perp) = \sqrt{2e}\,\frac{q_{\perp }}{M'}\,e^{-q_{\perp }^2/M'^2},
\end{equation}
as a result the Sivers function satisfies the positivity bound for all values of 
$x_a$ and $q_{\perp }$. If we  define the parameter
\begin{equation}
\rho = \frac{M'^2}{\langle q_\perp^2 \rangle +M'^2}\, ,
\end{equation}
such that $0< \rho < 1$, then  Eq.~(\ref{eq:siv-par-1}) becomes
\begin{equation}\label{eq:siv-par}
\Delta^N\! f_{a/p^\uparrow}(x_a,q_{\perp }) =   2 \,  \frac{\sqrt{2e}}{\pi}   \, {\cal N}_a(x_a)\,
f_{a/p}(x_a)\,\sqrt{\frac{1-\rho}{\rho}}\,q_{\perp }\,
\frac{e^{-q_{\perp }^2/ \rho\langle q_{\perp }^2 \rangle}}
{\langle q_{\perp }^2 \rangle^{3/2}}~.
\end{equation}
The unpolarized gluon TMD sitting in the denominator of the asymmetry  is parametrized as Gaussian distribution 
\begin{equation}
f_{a/p}(x_a,q_{\perp })  = \frac{1}{\pi \langle q_{\perp }^2 \rangle}
f(x_a)e^{-q_{\perp }^2/\langle q_{\perp }^2 \rangle}.
\end{equation}
The best fit parameters of Sivers function have been extracted for quarks \cite{Anselmino:2016uie} 
and gluons  \cite{DAlesio:2015fwo,DAlesio:2018rnv} from SIDIS  and RHIC data  respectively. In Ref. \cite{DAlesio:2018rnv}, 
new set of best fit parameters of GSF are extracted for $\langle q_{\perp }^2 \rangle=1$ GeV$^2$ and are 
tabulated in \tablename{~\ref{table1}}.

\begin{table}[H]
\centering
\begin{tabular}{ | >{\centering\arraybackslash}m{2cm}| >{\centering\arraybackslash}m{1.2cm}| 
>{\centering\arraybackslash}m{1.2cm}| >{\centering\arraybackslash}m{1.2cm}| 
>{\centering\arraybackslash}m{1.2cm}| >{\centering\arraybackslash}m{1.2cm}| 
>{\centering\arraybackslash}m{2cm}| >{\centering\arraybackslash}m{2cm}| 
>{\centering\arraybackslash}m{1.5cm}| }
\hline
\multicolumn{8}{ |c| }{Best fit parameters} \\
\cline{1-8}
Evolution & $a$ & $N_a$ & $\alpha$ & $\beta$ &$\rho$ & $\langle q^2_\perp\rangle$ 
GeV$^2$ & 
Notation  \\ \cline{1-8}
\multicolumn{1}{ |c  } {\multirow{3}{*}{DGLAP}} &
\multicolumn{1}{ |c| } {$g$ \cite{DAlesio:2015fwo}} & 0.65 & 2.8 & 2.8 & 0.687  & 0.25 & SIDIS1 \\
  \cline{2-8}
  \multicolumn{1}{ |c  }{} &
  \multicolumn{1}{ |c| } {$g$ \cite{DAlesio:2015fwo}} & 0.05 & 0.8 & 1.4 & 0.576  & 0.25 & SIDIS2 \\
  \cline{2-8}
  \multicolumn{1}{ |c  }{} &
  \multicolumn{1}{ |c| } {$g$ \cite{DAlesio:2018rnv}} & 0.25 & 0.6 & 0.6 & 0.1  & 1.0 & SIDIS3 \\
  \cline{1-8}
  \multicolumn{1}{ |c  } {\multirow{2}{*}{TMD}} &
  \multicolumn{1}{ |c| } {$u$ \cite{Echevarria:2014xaa}} & 0.106 & 1.051 & 4.857 &    & 0.38 & TMD-a \\
  \cline{2-7}
  \multicolumn{1}{ |c  }{} &
  \multicolumn{1}{ |c| } {$d$ \cite{Echevarria:2014xaa}} & -0.163 & 1.552 & 4.857 &    & 0.38 & TMD-b \\
  \cline{1-8}
\end{tabular}
\caption{\label{table1}Best fit parameters of Sivers function.}
\end{table}
%------------------------------------------------

%----------------------------------------------------------
\section{TMD Evolution}\label{sec3}
In DGLAP evolution, collinear PDFs evolve with only the probing scale ($\mu$). However, in TMD evolution 
approach, TMDs evolve with both the intrinsic transverse momentum ($p_{a\perp}$) of the parton and the probing scale. The TMD evolution
framework is derived in the impact parameter space ($b_{\perp}$) \cite{Aybat:2011zv}
\begin{equation}\label{ub1}
f(x_a,b_\perp,\mu)=\int d^2{\bm p}_{a \perp} 
e^{-i{\bm b}_\perp\cdot{\bm p}_{a\perp}}f(x_a,p_{a\perp},\mu), 
\end{equation}
and in the momentum space is given by 
\begin{equation}\label{ub2}
f(x_a,p_{a\perp},\mu)=\frac{1}{(2\pi)^2}\int d^2{\bm b}_\perp 
e^{i{\bm b}_\perp\cdot{\bm p}_{a\perp}}f(x_a,b_\perp,\mu).
\end{equation}

Usually, TMDs depend on two scales that are renormalization scale ($\mu$) and auxiliary  scale ($\zeta$) 
\cite{jcollins,aybat}.
In order to cure the light-cone (rapidity) divergences in TMD factorization, the scale $\zeta$ has been 
introduced. One can obtain the renormalization group  and Collins-Soper equations by taking
scale evolution with respect to $\mu$ and $\zeta$ respectively.
The unpolarized TMD expression at a given final scale $Q_f=\sqrt{\zeta}=M$ is obtained by solving RG and CS
equations \cite{Echevarria:2014xaa,jcollins,aybat} and is given below
\begin{eqnarray}\label{pert}
  f(x_a,b_{\perp},Q_f,\zeta)=f(x_a,b_\perp,Q_i)R_{pert}\left(Q_f,Q_i,b_{\ast}\right)
  R_{NP}\left(Q_f,b_{\perp}\right).
 \end{eqnarray}
where $Q_i=c/b_{\ast}(b_{\perp})$ is the initial scale of the TMD with 
$c=2e^{-\gamma_\epsilon}$ and $\gamma_\epsilon\approx0.577$. 
In line with Ref.\cite{Echevarria:2014xaa}, we adopt the $b_{\ast}$ prescription
  to avoid hitting the Landau pole by freezing the scale $b_{\perp}$. Here, 
 $b_{\ast}(b_{\perp})=\frac{b_{\perp}}{\sqrt{1+\left(\frac{b_{\perp}}{b_{\mathrm{max}}}\right)^2}}
 \approx b_{\mathrm{max}}$ when $b_{\perp}\rightarrow \infty$ and $b_{\ast}(b_{\perp})\approx
 b_{\perp}$ when $b_{\perp}\rightarrow 0$.
The $R_{pert}$  and $R_{NP}$
are the  perturbative and the nonperturbative parts of 
the TMD respectively, which are given below
 \be\label{sudakov}
 R_{pert}\left(Q_f,b_{\ast}\right)=\mathrm{exp}\Big\{{-\int_{c/b_{\ast}}^{Q_f}\frac{d\mu}
 {\mu}\left(A\log\left(\frac{Q_f^2} {\mu^2}\right)+B\right)}\Big\},
 \ee
 and
 \be
R_{NP}(Q_f,b_{\perp})=\mathrm{exp}\Bigg\{-\Big[g_1^{\mathrm{TMD}}+\frac{g_2}{2}
 \log\frac{Q_f}{Q_0}
 \Big]b_{\perp}^2\Bigg\},
\ee
where the anomalous dimensions are 
   $A=\sum_{n=1}^{\infty}\left(\frac{\alpha_s(\mu)}{\pi}\right)^nA_n$
 and $B=\sum_{n=1}^{\infty}\left(\frac{\alpha_s(\mu)}{\pi}\right)^nB_n$, and  
 the  coefficients for gluon case  are  $A_1=C_A$, 
$A_2=\frac12C_F\left(C_A\left(\frac{67}{18}-\frac{\pi^2}{6}\right)-\frac{5}{9}C_AN_f\right)$ 
 and $B_1=-\frac{1}{2}(\frac{11}{3}C_A-\frac{2}{3}N_f)$, and 
 for quark case  are $A_1=C_F$, $A_2=\frac12C_F\left(C_A\left(\frac{67}{18}-\frac{\pi^2}{6}\right)-
 \frac{5}{9}N_f\right)$ and $B_1=-\frac{3}{2}C_F$ \cite{Echevarria:2014xaa}.
 The unpolarized TMD at the initial scale can be written as 
 \be
  f(x_a,b_\perp,Q_i)=\sum_{i=g,q}\int_x^1\frac{d\hat{x}}{\hat{x}}
  C_{i/a}(x_a/\hat{x},b_{\perp},\alpha_s,Q_i)
  f_{i/p}(\hat{x},c/b_{\ast})+\mathcal{O}(b_{\perp}\varLambda_{QCD}),
  \ee
 where $C_{i/a}$ is the perturbatively calculated process independent coefficient function, and  is different for each type of TMD.  The derivative of the Sivers function follows the same 
evolution equation as given in Eq.\eqref{pert} but the $g_1^{\mathrm{TMD}}$  changes to 
$g_1^{\mathrm{Sivers}}$ in the $R_{NP}$ factor. The Sivers function $f_{1T}^{\perp }(x_a,{p}_{a\perp},Q_f)$
and it's derivative are related by 
Fourier transformation as below \cite{aybat} 
\be\label{ub5}
f_{1T}^{\perp }(x_a,{ p}_{a\perp},Q_f)=-\frac{1}{2\pi p_{a\perp}}\int_0^\infty db_\perp b_\perp 
J_1(p_{a\perp}b_\perp)f_{1T}^{\prime \perp }(x_a,b_\perp,Q_f),
\ee
and the unpolarized  TMD is given by
\be\label{ub5}
f_{a/p}(x_a, p_{a\perp},Q_f)=\frac{1}{2\pi}\int_0^\infty db_\perp b_\perp 
J_0({p}_{a\perp}b_\perp)f_{a/p}(x_a,b_\perp,Q_f),
\ee
The derivative of the Sivers function at the initial scale $Q_i$ can be  written in terms of Qiu-Sterman
function as below \cite{Boer:2003cm, Ji:2006ub}
\be\label{et6}
 f_{1T}^{\prime \perp  }(x_a,b_\perp,Q_i)\simeq\frac{M_pb_\perp}{2}T_{a,F}(x_a,x_a,Q_i),
 \ee
 where  $T_{a,F}(x_a,x_a,Q_i)$ is the Qiu-Sterman function, which is usually assumed to be 
proportional to collinear PDF \cite{Kouvaris:2006zy,Echevarria:2014xaa}
 \be
 T_{a,F}(x_a,x_a,Q_i)=\mathcal{N}_a(x_a)f_{a/p}(x_a,Q_i),
 \ee
the definition of $ \mathcal{N}_a(x_a)$  is given in Eq.\eqref{ngx}.  Note that here we have used the fact
 that Sivers function in Drell-Yan process have opposite sign compared to the Sivers function in semi-inclusive DIS. So far the best fit parameters of GSF  have not 
been extracted in the TMD evolution approach. However only the $u$ and $d$ quark Sivers function 
are known which were extracted in  \cite{Echevarria:2014xaa} from SIDIS data within TMD evolution scheme, which 
are tabulated in \tablename{~\ref{table1}}.
As so far a fit for the gluon Sivers function is not avavilable in the TMD evolution approach, in order to show the effect of this evolution on the asymmetry, we use an exploratory approach, namely, following Ref.\cite{Boer:2003tx},
 we define the following two set of parametrizations
for GSF by using the known $u$ and $d$
 quark Sivers function parameters
 \be \label{ab}
(a)~~\mathcal{N}_g (x_g) &=&(\mathcal{N}_u(x_g)+\mathcal{N}_d(x_g))/2 \nonumber\\
(b)~~\mathcal{N}_g (x_g)&=&\mathcal{N}_d(x_g).
\ee
We denote the first parametrization as TMD-a and second  one as TMD-b. As the sign of $N_u$ and $N_d$ are opposite, TMD-a gives a small GSF and TMD-b gives a large GSF.  As mentioned in the introduction, Burkardt sum rule 
\cite{ Burkardt:2004ur} gives a constraint on the GSF.  However, in order to implement the contstraint from the Burkardt sum rule contribution from all quark favours need to be included, and sea quark Sivers function is still not well constrained.  Assuming contributions only from $u$ and $d$ quarks and gluons, we have checked that TMD-a parametrization satisfies the Burkardt sum rule, violation is about $1\%$; wheras the TMD-b parametrization violates the sum rule by about $ 19\%$. The numerical values of best fit parameters are estimated \cite{Echevarria:2014xaa} at 
$Q_0=\sqrt{2.4}~\mathrm{GeV}$, $b_{\mathrm{max}}=1.5\mathrm{~GeV^{-1}}$, $g_2=0.16\mathrm{~GeV^2}$  and 
$\langle p^2_{s\perp }\rangle=0.282\mathrm{~GeV^2}$ with $g_1^{\mathrm{pdf}}=\langle p^2_{a\perp 
}\rangle/4=\langle q^2_{\perp 
}\rangle/4$ and $g_1^{\mathrm{sivers}}=\langle p^2_{s \perp  } \rangle/4$. The numerator and 
denominator parts of Eq.\eqref{eq:AN} in TMD evolution approach  can be written as the following
 \begin{align}\label{tmd2}
 \d\Delta\sigma  =-\frac{1}{\pi M_p}\frac{1}{2(2\pi)^2}\frac{1}{z(1-z)s}
 {\sum_a}\int \frac{dx_\gamma}{x_\gamma}  db_\perp b_\perp 
J_1(q_{\perp}b_\perp)f_{1T}^{\prime \perp }(x_a,b_\perp,Q_f)f_{\gamma/e}(x_\gamma)\frac{1}{2\hat{s}}
|\mathcal{M}_{\gamma a\rightarrow J/\psi a}|^2\sin(\phi_q),
\end{align}
\begin{equation}\label{tmd2}
 \begin{aligned}
2d\sigma
={}&\frac{1}{2(2\pi)^2\pi}\frac{1}{z(1-z)s}{\sum_a}\int \frac{dx_\gamma}{x_\gamma} db_\perp b_\perp 
J_0({q}_{\perp}b_\perp)f_{a/p}(x_a,b_\perp,Q_f)
f_{\gamma/e}(x_\gamma)\frac{1}{2\hat{s}}
|\mathcal{M}_{\gamma a\rightarrow J/\psi a}|^2.
\end{aligned}
\end{equation}
%========================================
\section{Numerical Results}\label{sec4}
In this section, we discuss the numerical results of Sivers asymmetry in 
$ep^{\uparrow}\to J/\psi+\mathrm{jet} +X$ photoproduction process, where the proton is transversely polarized. 
We consider the situation  $|{\bm q}_\perp|\ll|{\bm K}_\perp|$ i.e., the produced pair of $J/\psi$ and 
jet are almost back to back in the transverse plane as shown in \figurename{~\ref{fig1}}. This configuration
is feasible in the future proposed Electron-Ion collider (EIC) with C.M energy from 20 to 150 GeV. 
The NLO photon-gluon fusion and quark (anti-quark) initiated subprocesses, 
$\gamma g \rightarrow J/\psi g $ and $\gamma q (\mathrm{~or~}\bar{q})\rightarrow J/\psi q( 
\mathrm{~or~} \bar{q})$, are considered. The NRQCD model is 
employed for $J/\psi$ production, and the CS and CO states are considered for both numerator
and denominator parts of Eq.\eqref{eq:AN}. If we do not detect the jet in the final state then CS state
does not contribute to the asymmetry, because the initial and final state interactions between the final state
parton and remnant of the proton get canceled with each other as discussed in
\cite{Yuan:2008vn}. The values of long distance matrix elements (LDMEs) are taken from 
Ref.\cite{Chao:2012iv}. There are different set of LDMEs in the literature and the asymmetry is found to
be independent of the choice of LDME set.\par
There are two types of $J/\psi$ photoproductions that are resolved
 and direct photon contributions. The resolved photoproduction, the photon splits into partons which subsequently interact with 
 the partons of proton,  contributes to $J/\psi$ production in the low $z$ region ($z<0.3$).
 While in the direct photoproduction, the photon directly interacts electromagnetically with 
 the partons from the proton. For direct inelastic $J/\psi$ photoproduction one has to consider
 $0.3<z<0.9$ as discussed in Ref.\cite{Rajesh:2018qks}. The fragmentation of gluon and heavy quark can 
 also contribute to $J/\psi$ production at high transverse momentum of the  $J/\psi$ \cite{Li:1996jk}. The feed-down contribution 
from an excited state $\psi(2S)$ and 
the decay of $\chi_c$ states contribution to $J/\psi$ are $15\%$ \cite{Chekanov:2002at} and $1\%$ 
\cite{Butenschoen:2009zy,Artoisenet:2009xh} respectively, are not considered in this work. The final state 
parton becomes soft at $z\rightarrow 1$ which leads to infrared singularity. Therefore, to calculate the asymmetry
for direct inelastic $J/\psi$ photoproduction we consider $z=0.3$. The mass of 
$J/\psi$ is taken to be $M=3.1$ GeV.
The cteq6l1 PDF sets are used for collinear PDFs \cite{Buckley:2014ana}.\par
The Sivers asymmetry is calculated at EIC for $\sqrt{s}=45$  and 100 GeV within the DGLAP and TMD evolution
approaches. In DGLAP evolution approach, the GSF has been extracted in \cite{DAlesio:2015fwo}
from pion data at RHIC for fixed Gaussian width  $\langle q_{\perp }^2 \rangle=0.25$ GeV$^2$. Recently refitted 
the RHIC data with new set of GSF parameters for $\langle q_{\perp }^2 \rangle=1$ GeV$^2$ 
\cite{DAlesio:2018rnv}. However, GSF has not been  extracted yet in TMD evolution approach. The $u$ and $d$
quark Sivers functions are extracted in Ref.\cite{Echevarria:2014xaa} using TMD evolution approach. For 
numerical estimation of Sivers asymmetry, the $u$ and $d$ quark Sivers function best fit parameters are 
used for GSF as defined in Eq.\eqref{ab} wthin TMD evolution approach. The best fit parameters of GSF and quark Sivers function 
are tabulated in \tablename{~\ref{table1}}. The convention of figures as follows. The obtained 
Sivers asymmetry in DGLAP approach is represented with SIDIS1, SIDIS2 and SIDIS3. The TMD-a and TMD-b 
represent the Sivers asymmetry in TMD evolution approach.\par
\begin{figure}[H]
\begin{center}
\includegraphics[trim = 0.cm 0cm 1cm 0cm, width=8.5cm]{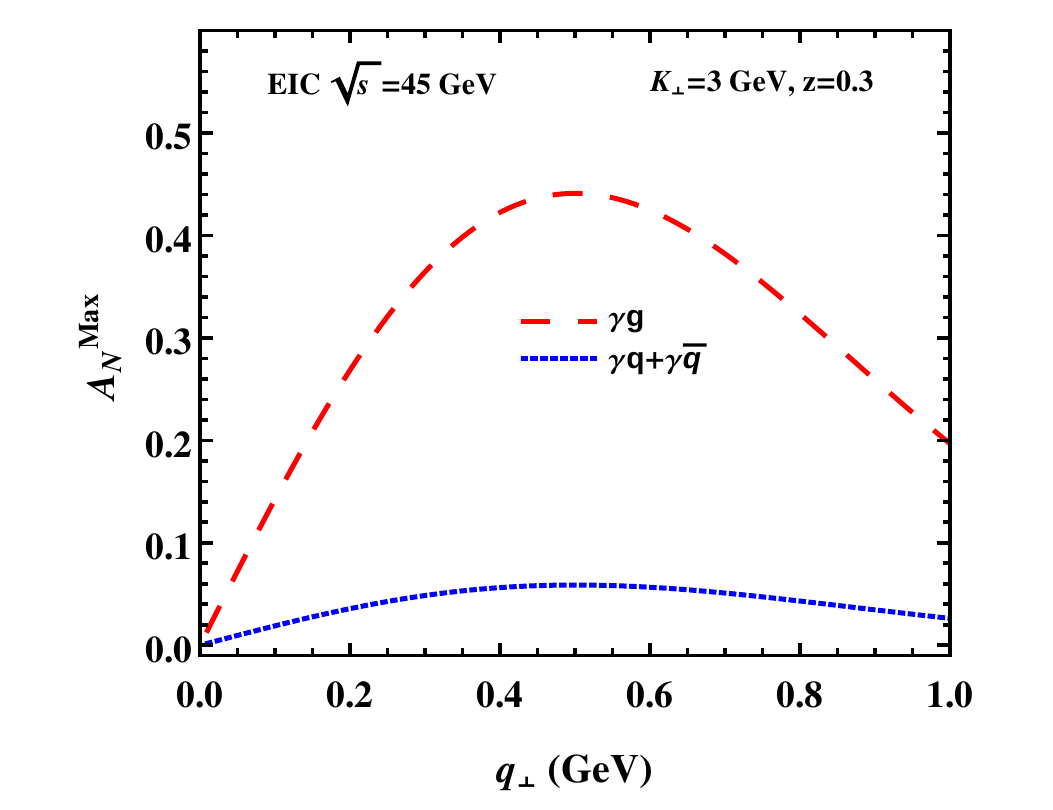}
\includegraphics[trim = 0.cm 0cm 1cm 0cm,width=8.5cm]{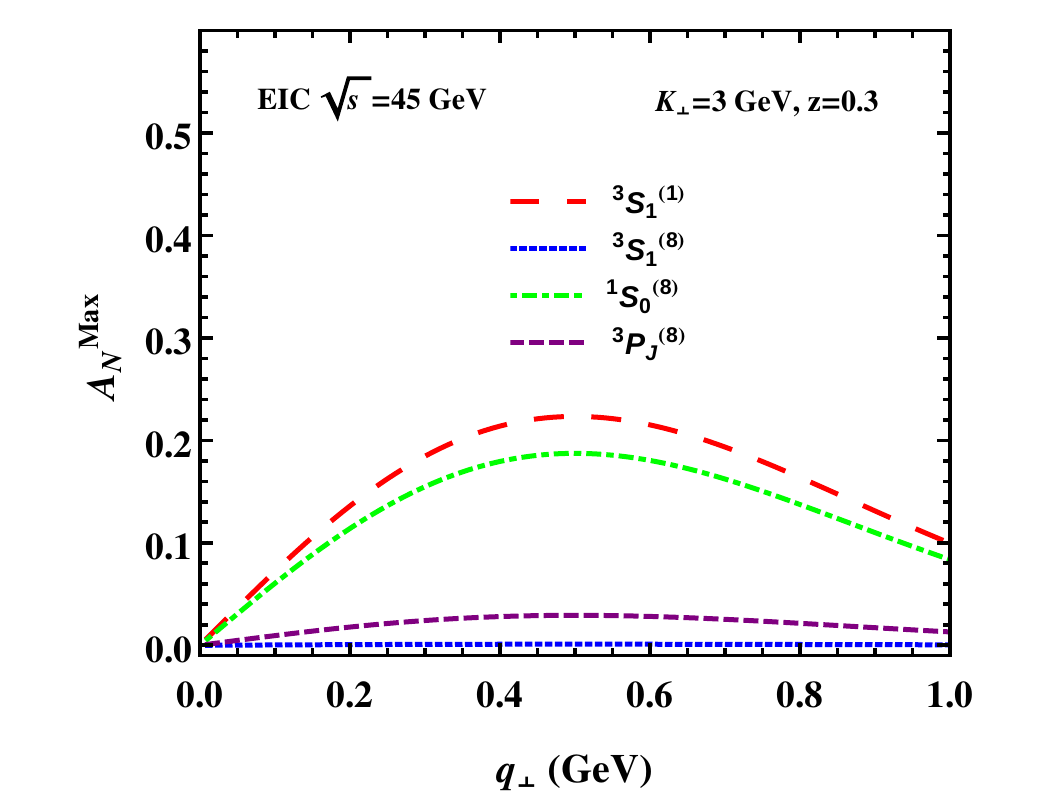}
\caption{(color online) Maximized Sivers asymmetry  in $e+p^{\uparrow}\to J/\psi+\mathrm{jet} +X$
 process as a function of  $q_\perp$ at EIC
  $\sqrt{s}=45$ GeV. 
  The Sivers function is saturated by adopting $\mathcal{N}_g(x)=1$ and $\rho=2/3$ for the 
  parametrization of Sivers function given in Eq.\eqref{eq:siv-par}.
  Left panel: for gluon and quark (antiquark) initiated subprocesses contribution to the asymmetry.
  Right panel: for different CS and CO  states 
  contribution to the maximum asymmetry.}
\label{fig2}
\end{center}
\end{figure}

In \figurename{~\ref{fig2}~}-{~\ref{fig4}}, the asymmetry in DGLAP  and TMD evolution approaches as a function of $q_\perp$ is shown
at $\sqrt{s}=45$ and 100 GeV for $K_\perp=3$  GeV respectively at  $z=0.3$. The asymmetry is shown in the 
range $0\leq q_\perp \leq 1$ GeV which is considered to satisfy the condition  $|{\bm q}_\perp|\ll|{\bm K}_\perp|$.
The value of $K_\perp=3$  GeV is chosen of the order of $J/\psi$ mass. For higher values of 
$K_\perp$ and $z$ the gluon channel contribution is suppressed because the momentum fraction of the parton, $x_a$,
depends quadratically on $K_\perp$ which can be seen from Eq.\eqref{delta}.\par
\begin{figure}[H]
\begin{center}
\includegraphics[trim = 0.cm 0cm 1cm 0cm, width=8.5cm]{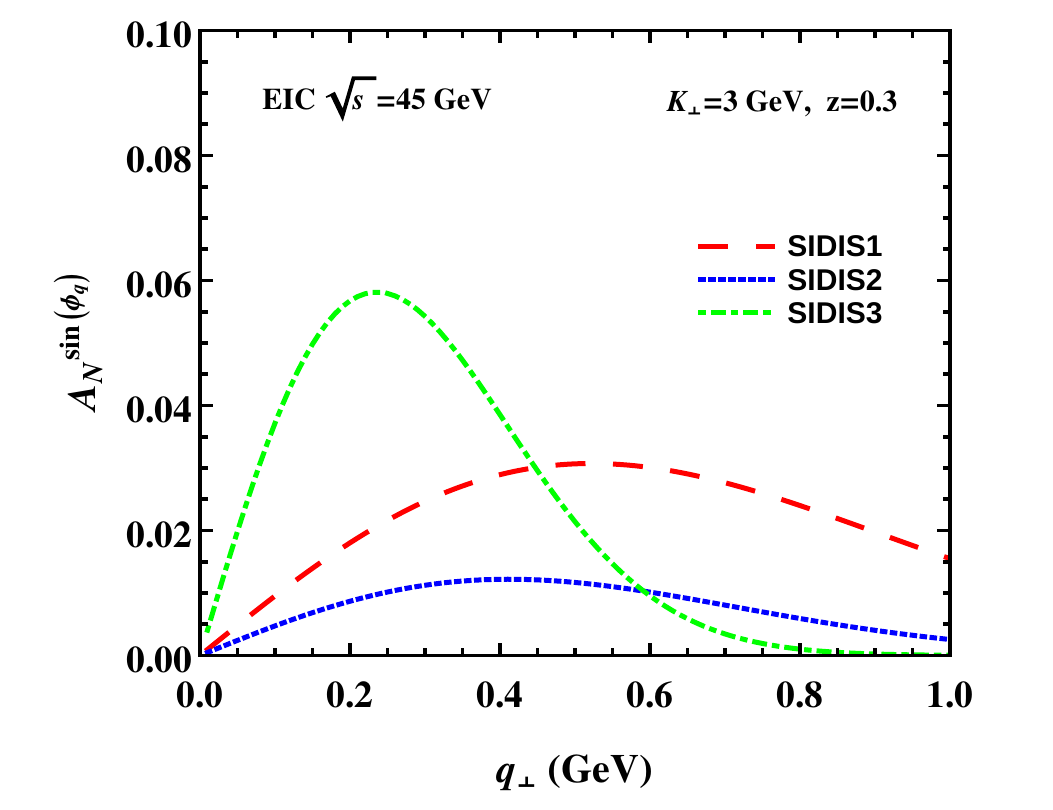}
\includegraphics[trim = 0.cm 0cm 1cm 0cm,width=8.5cm]{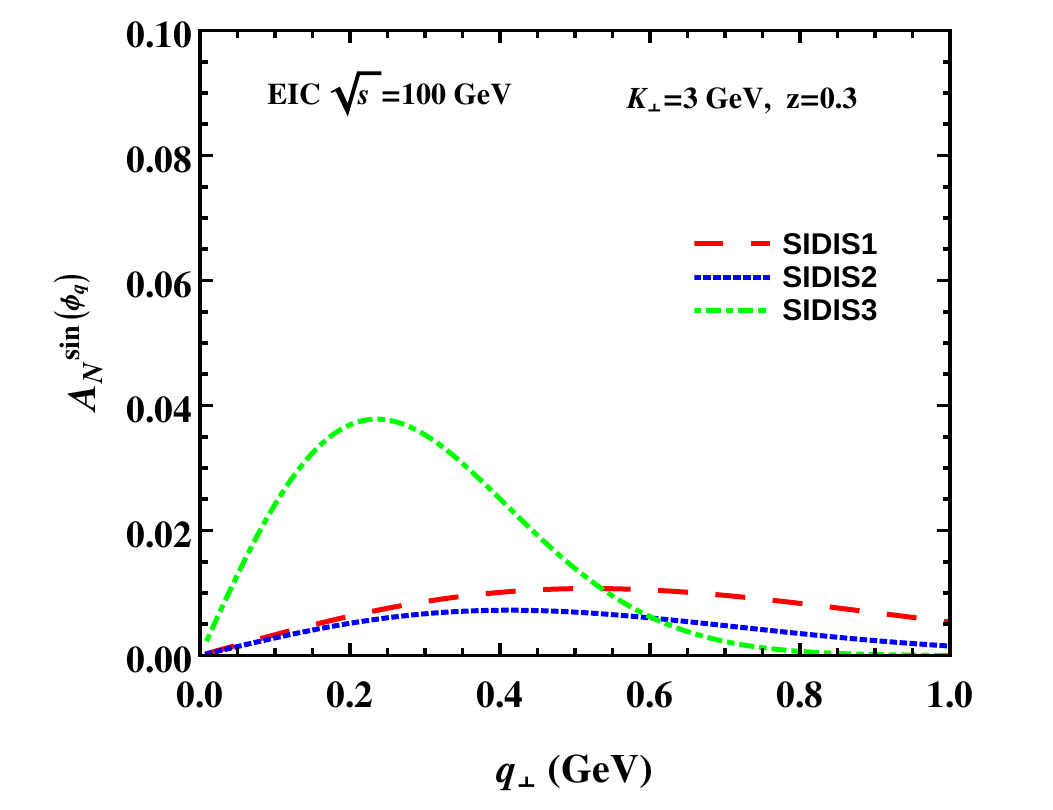}
\caption{(color online) The weighted Sivers asymmetry  in $e+p^{\uparrow}\to J/\psi+\mathrm{jet} +X$
 process as a function of  $q_\perp$ at EIC
 (a) $\sqrt{s}=45$ GeV (left panel) and  (b) $\sqrt{s}=100$ GeV (right panel) 
 using DGLAP evolution approach for SIDIS1, SIDIS2 and SIDIS3 GSF parametrization sets which are given in 
 \tablename{~\ref{table1}}.}
\label{fig3}
\end{center}
\end{figure}
\begin{figure}[H]
\begin{center}
\includegraphics[trim = 0.cm 0cm 1cm 0cm, width=8.5cm]{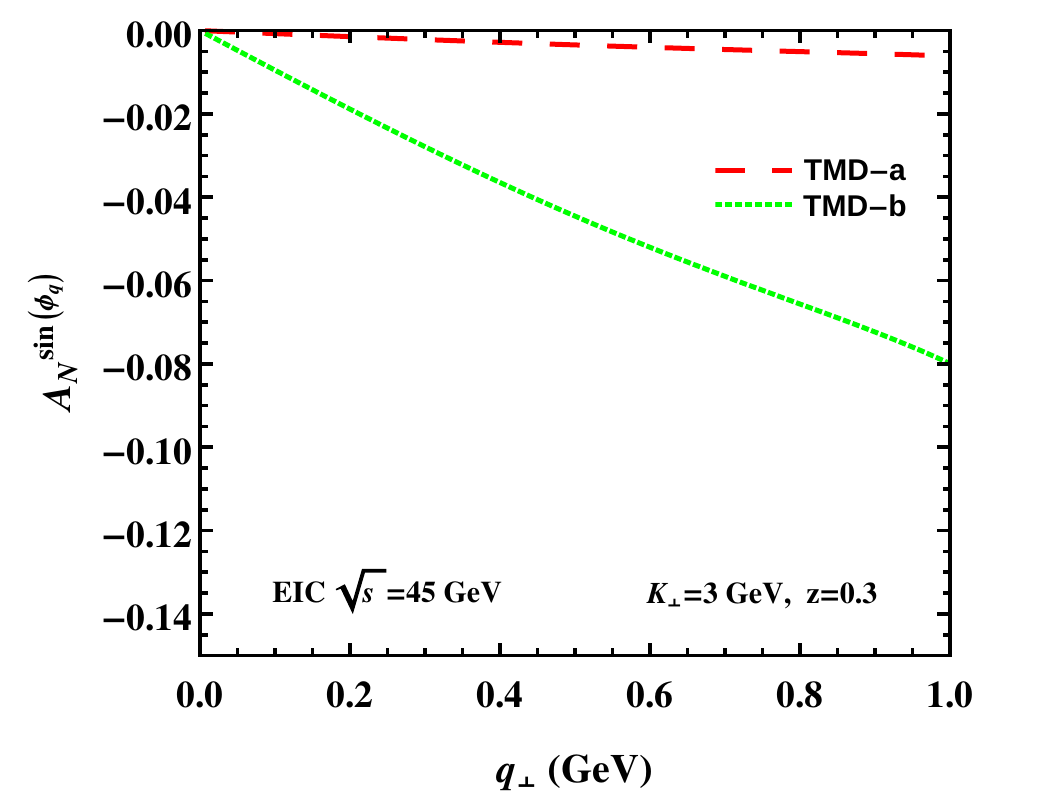}
\includegraphics[trim = 0.cm 0cm 1cm 0cm,width=8.5cm]{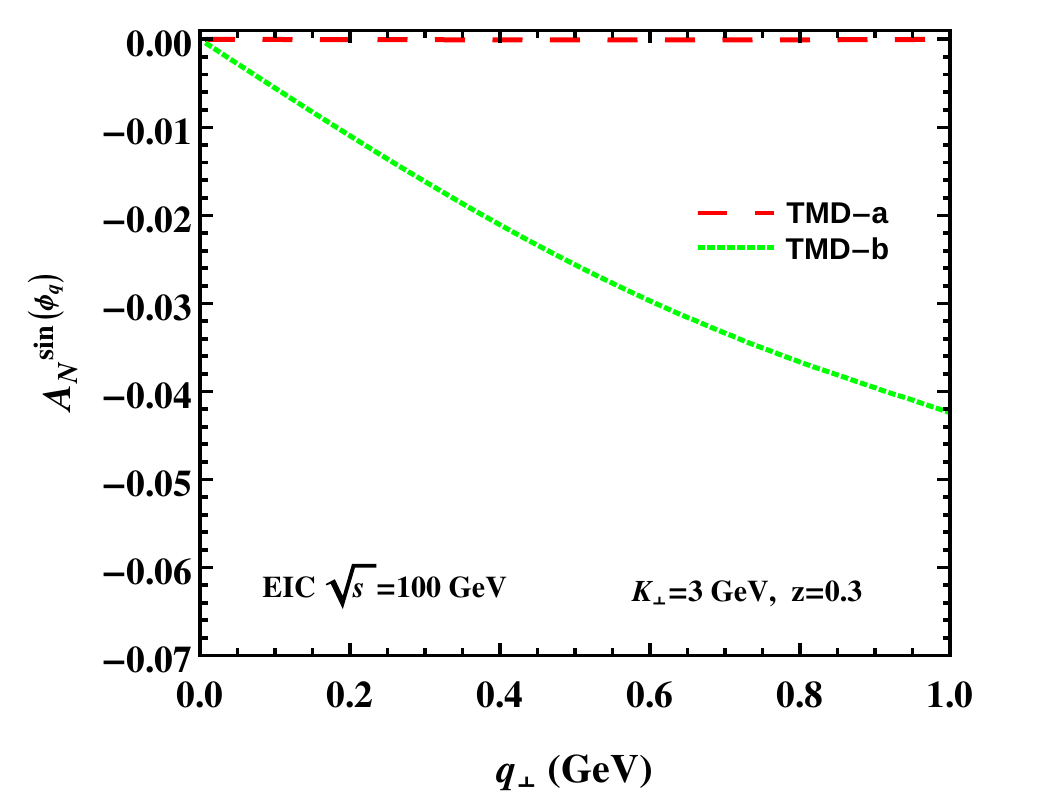}
\caption{ (color online) The weighted Sivers asymmetry  in $e+p^{\uparrow}\to J/\psi+\mathrm{jet} +X$
 process as a function of  $q_\perp$ at EIC
 (a) $\sqrt{s}=45$ GeV (left panel) and  (b) $\sqrt{s}=100$ GeV (right panel) 
 using TMD evolution approach for TMD-a  and TMD-b GSF parametrization sets which are given in 
 \tablename{~\ref{table1}}.}
\label{fig4}
\end{center}
\end{figure}

The maximized Sivers asymmetry, $A_N^{Max}$, at $\sqrt{s}=45$ GeV is shown in \figurename{\ref{fig2}}. Here, we saturated the Sivers function
bound by adopting $\mathcal{N}_a=1$ and $\rho=2/3$ \cite{DAlesio:2018rnv} in the parametrization of Sivers function which is given
in Eq.\eqref{eq:siv-par-1}. In the left panel of \figurename{\ref{fig2}}, the gluon and quark (antiquark)
channels contribution to $A_N^{Max}$ is shown, and the quark (antiquark) channel contribution is 
insignificant compared to gluon channel. On this basis we can say that the pair of 
$J/\psi$+jet photoproduction process is an effective channel that probes the poorly known GSF in the kinematical region considered here. We have neglected the quark channel contribution to
the numerator part of the asymmetry in  \figurename{~\ref{fig3}~}-{~\ref{fig4}}. In the right panel of 
\figurename{\ref{fig2}}, the individual CS and CO states contribution to $A_N^{Max}$ is shown.
The $\leftidx{^{3}}{S}{_1}^{(1)}$ and $\leftidx{^{1}}{S}{_0}^{(8)}$ states contribute largely to   $A_N^{Max}$ which is 
independent of $\sqrt{s}$.\par

 In the left panel of  \figurename{~\ref{fig3}~}, the weighted Sivers asymmetry, $A_N^{\sin(\phi_q)}$, is estimated to be about 3\%,
1\% and 6\% respectively for SIDIS1, SIDIS2 and SIDIS3 set of GSF parameters at $\sqrt{s}=45$ GeV.
The $A_N^{\sin(\phi_q)}$ is reduced about 2\% for  $\sqrt{s}=$100 GeV as shown in the right panel of 
\figurename{~\ref{fig3}~}. In \figurename{~\ref{fig4}}, negative Sivers asymmetry is shown in TMD evolution
approach.
The sign of the Sivers asymmetry depends on relative magnitude of $N_u$ and $N_d$ and these have opposite sign which
can be observed in \tablename{ \ref{table1}}.
 For TMD-b parameter set, the $\mathcal{N}_g$ is assumed to be proportional to $\mathcal{N}_d$ of 
 $d$ quark, see the Eq.\eqref{ab}. The values of $N_d$ is negative which leads to 
negative asymmetry.  The average of $u$ and $d$ quarks $x_a$-dependent factor, $\mathcal{N}_a$, 
is defined for gluon as given in Eq.\eqref{ab} for TMD-a parameter set. 
The magnitude of $\mathcal{N}_d$ is comparable but slightly dominant compared to 
$\mathcal{N}_u$ that leads to negative and small asymmetry for TMD-a parameter set.
In \figurename{~\ref{fig4}}, $A_N^{\sin(\phi_q)}$ is estimated maximum of 8\% and 4\% at $\sqrt{s}=45$ and 100 GeV 
for TMD-b parameter set.

%---------------------------------------


\section{Conclusion}\label{sec5}

In this work, we gave an estimate of the Sivers asymmetry in almost
back-to-back $J/\psi$ and jet photoproduction at the future EIC. We assumed
TMD factorization for this process and used generalized parton model,
incorporating the intrinsic transverse momenta. The quasi-real
photoproduction takes place through the  Weizs$\ddot{a}$ker-Williams photon
distribution of the electron. We used NRQCD to calculate the $J/\psi$ 
production and incorporated both CS and CO contributions to the asymmetry.
Major contribution comes from $^3S_1^{(1)}$ and $^1S_0^{(8)}$ states. We have
also shown the effect of the TMD evolution on the asymmetry. In fact the
Sivers asymmetry is positive without incorporating the TMD evolution,
whereas it becomes negative when evolution is incorporated. We have obtained
sizable Sivers asymmetry where the main contribution comes from the gluon
Sivers function and the quark contribution is small. Therefore, back-to-back
production of $J/\psi$ and jet at the future EIC is a promising tool to
access the gluon Sivers function.

\section*{Acknowledgment}
We would like to thank Cristian Pisano and Pieter Taels for useful
discussions. The work of S.R. is 
supported by  Fondazione Sardegna under the project “Quarkonium
at LHC energies”, CUP F71I17000160002 (University of Cagliari).
A.M. would like to thank University of Cagliari and INFN, Cagliari, Italy
for hospitality where the final stage of this work was completed.

\appendix
\section*{Appendices}
 \section{K\lowercase{inematics}}\label{ap1}
 We consider the frame in which the proton and electron are moving along +$z$ and -$z$-axises respectively and 
their four momenta are given by
\be
P=\frac{\sqrt{s}}{2}(1,0,0,1),~~l=\frac{\sqrt{s}}{2}(1,0,0,-1).
\ee
The C.M energy of electron-proton system is $s=(P+l)^2$.
The above four momenta in light-cone coordinate system can be written as 
\be
P^\mu=\sqrt{\frac s2}n_+^\mu,~~ l^\mu=\sqrt{\frac s2}n_-^\mu,
\ee
where $n_+$ and $n_-$ are two light-like vectors with $n_+\cdot n_-=1$ and $n_+^2=n^2_-=0$. 
\be
n_+^\mu=(1,0,{\bm 0}),~~~n_-^\mu=(0,1,{\bm 0}).
\ee
We assume that the quasi-real photon is collinear to the electron.
The quasi-real photon and parton four momenta are given by
\be
q^\mu=x_\gamma\sqrt{\frac s2}n_-^\mu,
\ee
\be
p=\frac{p^2_{\perp a}}{2x_a\sqrt{\frac s2}}n_-^\mu  +x_a\sqrt{\frac s2}n_+^\mu + {\bm p}^\mu_{\perp a} \approx
x_a\sqrt{\frac s2}n_+^\mu + {\bm p}^\mu_{\perp a} ,
\ee
where $x_\gamma=\frac{q^-}{l^-}$ and $x_a=\frac{p^+}{P^+}$ are the light-cone momentum fractions. 
 The four momentum of the $J/\psi$ and final parton are given by
\be
P_\Psi^\mu=zx_\gamma\sqrt{\frac s2}n_-^\mu+\frac{M^2+P^2_{\Psi\perp}}{2zx_\gamma\sqrt{\frac s2}}n_+^\mu+{\bm P}_
{\Psi \perp}^\mu.
\ee
\be
P_j^\mu=z_1x_\gamma\sqrt{\frac s2}n_-^\mu+\frac{P^2_{j\perp}}{2z_1x_\gamma\sqrt{\frac s2}}n_+^\mu+{\bm P}_{j\perp}^\mu.
\ee
The inelastic variables are defined as $z=\frac{P\cdot P_\Psi}{P\cdot q}=\frac{P_\Psi^-}{q^-}$ and $z_1=\frac{P\cdot P_j}{P\cdot q}=\frac{P_j^-}{q^-}$.
By using the above relations, we can write down the expressions of  Mandelstam variables as below
\be\label{sh}
\hat{s}=(q+p)^2=2k\cdot q=sx_ax_\gamma,
\ee
\be\label{th}
\hat{t}&=&(q-P_j)^2=-2q\cdot P_j=-\frac{P^2_{j\perp}}{z_1},
\ee
\be\label{uh}
\hat{u}&=&(q-P_\Psi)^2=M^2-2q\cdot P_\Psi\nonumber\\
&=&M^2-\frac{M^2+P^2_{\Psi \perp}}{z}.
\ee
Here $M$ being the mass of  $J/\psi$. \\

%===============================================
\section{M\lowercase{atrix} \lowercase{elements} \lowercase{for} $\gamma +q (\bar{q})\to J/\psi+ q(\bar{q}) $  \lowercase{subprocess}}\label{ap2}
In this section the matrix elements for $\gamma +q(\bar{q})\to J/\psi +q (\bar{q})$ channel are presented:
\be
|\mathcal{M}(^3S_1^{(1)})|^2=0,
\ee

\be
|\mathcal{M}(^3S_1^{(8)})|^2=\frac{-2(4\pi)^3e_c^2\alpha_s^2\alpha}{9M^3\hat{s}\hat{t}}
\langle 0|\mathcal{O}_8^{J/\psi}(^{3}S_1)|0\rangle\left[\hat{s}^2+\hat{t}^2+2\hat{u}M^2\right],
\ee

\be
|\mathcal{M}(^1S_0^{(8)})|^2=\frac{-4(4\pi)^3e_c^2\alpha_s^2\alpha}{3M}\langle 0|\mathcal{O}_8^{J/\psi}
(^{1}S_0)|0\rangle\frac{\hat{s}^2+\hat{t}^2}{\hat{u}(\hat{s}+\hat{t})^2},
\ee

\be
|\mathcal{M}(^3P_0^{(8)})|^2=\frac{-16(4\pi)^3e_c^2\alpha_s^2\alpha}{9M^3}\langle 0|\mathcal{O}
_8^{J/\psi}(^{3}P_0)|0\rangle\frac{(\hat{s}^2+\hat{t}^2)(\hat{u}-3M^2)^2}{\hat{u}(\hat{s}+\hat{t})^4},
\ee

\be
|\mathcal{M}(^3P_1^{(8)})|^2=\frac{-32 (4\pi)^3e_c^2\alpha_s^2\alpha}{9M^3}\langle
0|\mathcal{O}_8^{J/\psi}(^{3}P_1)|0\rangle\frac{(\hat{s}^2+\hat{t}^2)\hat{u}+4M^2\hat{s}\hat{t}}
{(\hat{s}+\hat{t})^4},
\ee

\be
|\mathcal{M}(^3P_2^{(8)})|^2=\frac{32(4\pi)^3e_c^2\alpha_s^2\alpha}{45 M^3  \hat{u} (\hat{s}+\hat{t})^4}\langle 0|\mathcal{O}_8^{J/\psi}(^{3}P_2)|0\rangle
\Big[-\hat{u}^2 \left(7 \hat{s}^2+12 \hat{s} \hat{t}+7 \hat{t}^2\right)\nonumber\\
-12 \hat{u} \left(\hat{s}^2+\hat{s}
\hat{t}+\hat{t}^2\right) (\hat{s}+\hat{t})
-6   \left(\hat{s}^2+\hat{t}^2\right) (\hat{s}+\hat{t})^2\Big].
\ee
 The matrix elements for $\gamma g\to J/\psi+ g $ channel are
given in our previous paper \cite{Rajesh:2018qks}.
%=======================================
\bibliographystyle{apsrev}
\bibliography{references}

\begin{thebibliography}{68}
\expandafter\ifx\csname natexlab\endcsname\relax\def\natexlab#1{#1}\fi
\expandafter\ifx\csname bibnamefont\endcsname\relax
  \def\bibnamefont#1{#1}\fi
\expandafter\ifx\csname bibfnamefont\endcsname\relax
  \def\bibfnamefont#1{#1}\fi
\expandafter\ifx\csname citenamefont\endcsname\relax
  \def\citenamefont#1{#1}\fi
\expandafter\ifx\csname url\endcsname\relax
  \def\url#1{\texttt{#1}}\fi
\expandafter\ifx\csname urlprefix\endcsname\relax\def\urlprefix{URL }\fi
\providecommand{\bibinfo}[2]{#2}
\providecommand{\eprint}[2][]{\url{#2}}

\bibitem[{\citenamefont{Sivers}(1990)}]{Sivers:1989cc}
\bibinfo{author}{\bibfnamefont{D.~W.} \bibnamefont{Sivers}},
  \bibinfo{journal}{Phys. Rev.} \textbf{\bibinfo{volume}{D41}},
  \bibinfo{pages}{83} (\bibinfo{year}{1990}).

\bibitem[{\citenamefont{Burkardt}(2004{\natexlab{a}})}]{Burkardt:2003uw}
\bibinfo{author}{\bibfnamefont{M.}~\bibnamefont{Burkardt}},
  \bibinfo{journal}{Nucl. Phys.} \textbf{\bibinfo{volume}{A735}},
  \bibinfo{pages}{185} (\bibinfo{year}{2004}{\natexlab{a}}),
  \eprint{hep-ph/0302144}.

\bibitem[{\citenamefont{Qiu and Sterman}(1991)}]{Qiu:1991pp}
\bibinfo{author}{\bibfnamefont{J.-w.} \bibnamefont{Qiu}} \bibnamefont{and}
  \bibinfo{author}{\bibfnamefont{G.~F.} \bibnamefont{Sterman}},
  \bibinfo{journal}{Phys. Rev. Lett.} \textbf{\bibinfo{volume}{67}},
  \bibinfo{pages}{2264} (\bibinfo{year}{1991}).

\bibitem[{\citenamefont{Airapetian et~al.}(2005)}]{Airapetian:2004tw}
\bibinfo{author}{\bibfnamefont{A.}~\bibnamefont{Airapetian}}
  \bibnamefont{et~al.} (\bibinfo{collaboration}{HERMES}),
  \bibinfo{journal}{Phys. Rev. Lett.} \textbf{\bibinfo{volume}{94}},
  \bibinfo{pages}{012002} (\bibinfo{year}{2005}), \eprint{hep-ex/0408013}.

\bibitem[{\citenamefont{Adolph et~al.}(2012)}]{Adolph:2012sp}
\bibinfo{author}{\bibfnamefont{C.}~\bibnamefont{Adolph}} \bibnamefont{et~al.}
  (\bibinfo{collaboration}{COMPASS}), \bibinfo{journal}{Phys. Lett.}
  \textbf{\bibinfo{volume}{B717}}, \bibinfo{pages}{383} (\bibinfo{year}{2012}),
  \eprint{1205.5122}.

\bibitem[{\citenamefont{Anselmino et~al.}(2017)\citenamefont{Anselmino,
  Boglione, D'Alesio, Murgia, and Prokudin}}]{Anselmino:2016uie}
\bibinfo{author}{\bibfnamefont{M.}~\bibnamefont{Anselmino}},
  \bibinfo{author}{\bibfnamefont{M.}~\bibnamefont{Boglione}},
  \bibinfo{author}{\bibfnamefont{U.}~\bibnamefont{D'Alesio}},
  \bibinfo{author}{\bibfnamefont{F.}~\bibnamefont{Murgia}}, \bibnamefont{and}
  \bibinfo{author}{\bibfnamefont{A.}~\bibnamefont{Prokudin}},
  \bibinfo{journal}{JHEP} \textbf{\bibinfo{volume}{04}}, \bibinfo{pages}{046}
  (\bibinfo{year}{2017}), \eprint{1612.06413}.

\bibitem[{\citenamefont{D'Alesio et~al.}(2015)\citenamefont{D'Alesio, Murgia,
  and Pisano}}]{DAlesio:2015fwo}
\bibinfo{author}{\bibfnamefont{U.}~\bibnamefont{D'Alesio}},
  \bibinfo{author}{\bibfnamefont{F.}~\bibnamefont{Murgia}}, \bibnamefont{and}
  \bibinfo{author}{\bibfnamefont{C.}~\bibnamefont{Pisano}},
  \bibinfo{journal}{JHEP} \textbf{\bibinfo{volume}{09}}, \bibinfo{pages}{119}
  (\bibinfo{year}{2015}), \eprint{1506.03078}.

\bibitem[{\citenamefont{D'Alesio
  et~al.}(2019{\natexlab{a}})\citenamefont{D'Alesio, Flore, Murgia, Pisano, and
  Taels}}]{DAlesio:2018rnv}
\bibinfo{author}{\bibfnamefont{U.}~\bibnamefont{D'Alesio}},
  \bibinfo{author}{\bibfnamefont{C.}~\bibnamefont{Flore}},
  \bibinfo{author}{\bibfnamefont{F.}~\bibnamefont{Murgia}},
  \bibinfo{author}{\bibfnamefont{C.}~\bibnamefont{Pisano}}, \bibnamefont{and}
  \bibinfo{author}{\bibfnamefont{P.}~\bibnamefont{Taels}},
  \bibinfo{journal}{Phys. Rev.} \textbf{\bibinfo{volume}{D99}},
  \bibinfo{pages}{036013} (\bibinfo{year}{2019}{\natexlab{a}}),
  \eprint{1811.02970}.

\bibitem[{\citenamefont{Aybat et~al.}(2012{\natexlab{a}})\citenamefont{Aybat,
  Collins, Qiu, and Rogers}}]{Aybat:2011ge}
\bibinfo{author}{\bibfnamefont{S.~M.} \bibnamefont{Aybat}},
  \bibinfo{author}{\bibfnamefont{J.~C.} \bibnamefont{Collins}},
  \bibinfo{author}{\bibfnamefont{J.-W.} \bibnamefont{Qiu}}, \bibnamefont{and}
  \bibinfo{author}{\bibfnamefont{T.~C.} \bibnamefont{Rogers}},
  \bibinfo{journal}{Phys. Rev.} \textbf{\bibinfo{volume}{D85}},
  \bibinfo{pages}{034043} (\bibinfo{year}{2012}{\natexlab{a}}),
  \eprint{1110.6428}.

\bibitem[{\citenamefont{Aybat and Rogers}(2011)}]{Aybat:2011zv}
\bibinfo{author}{\bibfnamefont{S.~M.} \bibnamefont{Aybat}} \bibnamefont{and}
  \bibinfo{author}{\bibfnamefont{T.~C.} \bibnamefont{Rogers}},
  \bibinfo{journal}{Phys. Rev.} \textbf{\bibinfo{volume}{D83}},
  \bibinfo{pages}{114042} (\bibinfo{year}{2011}), \eprint{1101.5057}.

\bibitem[{\citenamefont{Collins and Rogers}(2017)}]{Collins:2017oxh}
\bibinfo{author}{\bibfnamefont{J.}~\bibnamefont{Collins}} \bibnamefont{and}
  \bibinfo{author}{\bibfnamefont{T.~C.} \bibnamefont{Rogers}},
  \bibinfo{journal}{Phys. Rev.} \textbf{\bibinfo{volume}{D96}},
  \bibinfo{pages}{054011} (\bibinfo{year}{2017}), \eprint{1705.07167}.

\bibitem[{\citenamefont{Echevarria et~al.}(2016)\citenamefont{Echevarria,
  Scimemi, and Vladimirov}}]{Echevarria:2016scs}
\bibinfo{author}{\bibfnamefont{M.~G.} \bibnamefont{Echevarria}},
  \bibinfo{author}{\bibfnamefont{I.}~\bibnamefont{Scimemi}}, \bibnamefont{and}
  \bibinfo{author}{\bibfnamefont{A.}~\bibnamefont{Vladimirov}},
  \bibinfo{journal}{JHEP} \textbf{\bibinfo{volume}{09}}, \bibinfo{pages}{004}
  (\bibinfo{year}{2016}), \eprint{1604.07869}.

\bibitem[{\citenamefont{Echevarria et~al.}(2014)\citenamefont{Echevarria,
  Idilbi, Kang, and Vitev}}]{Echevarria:2014xaa}
\bibinfo{author}{\bibfnamefont{M.~G.} \bibnamefont{Echevarria}},
  \bibinfo{author}{\bibfnamefont{A.}~\bibnamefont{Idilbi}},
  \bibinfo{author}{\bibfnamefont{Z.-B.} \bibnamefont{Kang}}, \bibnamefont{and}
  \bibinfo{author}{\bibfnamefont{I.}~\bibnamefont{Vitev}},
  \bibinfo{journal}{Phys. Rev.} \textbf{\bibinfo{volume}{D89}},
  \bibinfo{pages}{074013} (\bibinfo{year}{2014}), \eprint{1401.5078}.

\bibitem[{\citenamefont{Accardi et~al.}(2016)}]{Accardi:2012qut}
\bibinfo{author}{\bibfnamefont{A.}~\bibnamefont{Accardi}} \bibnamefont{et~al.},
  \bibinfo{journal}{Eur. Phys. J.} \textbf{\bibinfo{volume}{A52}},
  \bibinfo{pages}{268} (\bibinfo{year}{2016}), \eprint{1212.1701}.

\bibitem[{\citenamefont{Brodsky et~al.}(2013)\citenamefont{Brodsky, Fleuret,
  Hadjidakis, and Lansberg}}]{Brodsky:2012vg}
\bibinfo{author}{\bibfnamefont{S.~J.} \bibnamefont{Brodsky}},
  \bibinfo{author}{\bibfnamefont{F.}~\bibnamefont{Fleuret}},
  \bibinfo{author}{\bibfnamefont{C.}~\bibnamefont{Hadjidakis}},
  \bibnamefont{and} \bibinfo{author}{\bibfnamefont{J.~P.}
  \bibnamefont{Lansberg}}, \bibinfo{journal}{Phys. Rept.}
  \textbf{\bibinfo{volume}{522}}, \bibinfo{pages}{239} (\bibinfo{year}{2013}),
  \eprint{1202.6585}.

\bibitem[{\citenamefont{Kikoła et~al.}(2017)\citenamefont{Kikoła, Echevarria,
  Hadjidakis, Lansberg, Lorcé, Massacrier, Quintans, Signori, and
  Trzeciak}}]{Kikola:2017hnp}
\bibinfo{author}{\bibfnamefont{D.}~\bibnamefont{Kikoła}},
  \bibinfo{author}{\bibfnamefont{M.~G.} \bibnamefont{Echevarria}},
  \bibinfo{author}{\bibfnamefont{C.}~\bibnamefont{Hadjidakis}},
  \bibinfo{author}{\bibfnamefont{J.-P.} \bibnamefont{Lansberg}},
  \bibinfo{author}{\bibfnamefont{C.}~\bibnamefont{Lorcé}},
  \bibinfo{author}{\bibfnamefont{L.}~\bibnamefont{Massacrier}},
  \bibinfo{author}{\bibfnamefont{C.~M.} \bibnamefont{Quintans}},
  \bibinfo{author}{\bibfnamefont{A.}~\bibnamefont{Signori}}, \bibnamefont{and}
  \bibinfo{author}{\bibfnamefont{B.}~\bibnamefont{Trzeciak}},
  \bibinfo{journal}{Few Body Syst.} \textbf{\bibinfo{volume}{58}},
  \bibinfo{pages}{139} (\bibinfo{year}{2017}), \eprint{1702.01546}.

\bibitem[{\citenamefont{Trzeciak et~al.}(2017)\citenamefont{Trzeciak, Da~Silva,
  Ferreiro, Hadjidakis, Kikola, Lansberg, Massacrier, Seixas, Uras, and
  Yang}}]{Trzeciak:2017csa}
\bibinfo{author}{\bibfnamefont{B.}~\bibnamefont{Trzeciak}},
  \bibinfo{author}{\bibfnamefont{C.}~\bibnamefont{Da~Silva}},
  \bibinfo{author}{\bibfnamefont{E.~G.} \bibnamefont{Ferreiro}},
  \bibinfo{author}{\bibfnamefont{C.}~\bibnamefont{Hadjidakis}},
  \bibinfo{author}{\bibfnamefont{D.}~\bibnamefont{Kikola}},
  \bibinfo{author}{\bibfnamefont{J.~P.} \bibnamefont{Lansberg}},
  \bibinfo{author}{\bibfnamefont{L.}~\bibnamefont{Massacrier}},
  \bibinfo{author}{\bibfnamefont{J.}~\bibnamefont{Seixas}},
  \bibinfo{author}{\bibfnamefont{A.}~\bibnamefont{Uras}}, \bibnamefont{and}
  \bibinfo{author}{\bibfnamefont{Z.}~\bibnamefont{Yang}}, \bibinfo{journal}{Few
  Body Syst.} \textbf{\bibinfo{volume}{58}}, \bibinfo{pages}{148}
  (\bibinfo{year}{2017}), \eprint{1703.03726}.

\bibitem[{\citenamefont{Mulders and Rodrigues}(2001)}]{Mulders:2000sh}
\bibinfo{author}{\bibfnamefont{P.~J.} \bibnamefont{Mulders}} \bibnamefont{and}
  \bibinfo{author}{\bibfnamefont{J.}~\bibnamefont{Rodrigues}},
  \bibinfo{journal}{Phys. Rev.} \textbf{\bibinfo{volume}{D63}},
  \bibinfo{pages}{094021} (\bibinfo{year}{2001}), \eprint{hep-ph/0009343}.

\bibitem[{\citenamefont{Burkardt}(2004{\natexlab{b}})}]{Burkardt:2004ur}
\bibinfo{author}{\bibfnamefont{M.}~\bibnamefont{Burkardt}},
  \bibinfo{journal}{Phys. Rev.} \textbf{\bibinfo{volume}{D69}},
  \bibinfo{pages}{091501} (\bibinfo{year}{2004}{\natexlab{b}}),
  \eprint{hep-ph/0402014}.

\bibitem[{\citenamefont{Anselmino et~al.}(2009)\citenamefont{Anselmino,
  Boglione, D'Alesio, Kotzinian, Melis, Murgia, Prokudin, and
  Turk}}]{Anselmino:2008sga}
\bibinfo{author}{\bibfnamefont{M.}~\bibnamefont{Anselmino}},
  \bibinfo{author}{\bibfnamefont{M.}~\bibnamefont{Boglione}},
  \bibinfo{author}{\bibfnamefont{U.}~\bibnamefont{D'Alesio}},
  \bibinfo{author}{\bibfnamefont{A.}~\bibnamefont{Kotzinian}},
  \bibinfo{author}{\bibfnamefont{S.}~\bibnamefont{Melis}},
  \bibinfo{author}{\bibfnamefont{F.}~\bibnamefont{Murgia}},
  \bibinfo{author}{\bibfnamefont{A.}~\bibnamefont{Prokudin}}, \bibnamefont{and}
  \bibinfo{author}{\bibfnamefont{C.}~\bibnamefont{Turk}},
  \bibinfo{journal}{Eur. Phys. J.} \textbf{\bibinfo{volume}{A39}},
  \bibinfo{pages}{89} (\bibinfo{year}{2009}), \eprint{0805.2677}.

\bibitem[{\citenamefont{Brodsky et~al.}(2002)\citenamefont{Brodsky, Hwang, and
  Schmidt}}]{Brodsky:2002rv}
\bibinfo{author}{\bibfnamefont{S.~J.} \bibnamefont{Brodsky}},
  \bibinfo{author}{\bibfnamefont{D.~S.} \bibnamefont{Hwang}}, \bibnamefont{and}
  \bibinfo{author}{\bibfnamefont{I.}~\bibnamefont{Schmidt}},
  \bibinfo{journal}{Nucl. Phys.} \textbf{\bibinfo{volume}{B642}},
  \bibinfo{pages}{344} (\bibinfo{year}{2002}), \eprint{hep-ph/0206259}.

\bibitem[{\citenamefont{Boer et~al.}(2003)\citenamefont{Boer, Mulders, and
  Pijlman}}]{Boer:2003cm}
\bibinfo{author}{\bibfnamefont{D.}~\bibnamefont{Boer}},
  \bibinfo{author}{\bibfnamefont{P.~J.} \bibnamefont{Mulders}},
  \bibnamefont{and} \bibinfo{author}{\bibfnamefont{F.}~\bibnamefont{Pijlman}},
  \bibinfo{journal}{Nucl. Phys.} \textbf{\bibinfo{volume}{B667}},
  \bibinfo{pages}{201} (\bibinfo{year}{2003}), \eprint{hep-ph/0303034}.

\bibitem[{\citenamefont{Adamczyk et~al.}(2016)}]{Adamczyk:2015gyk}
\bibinfo{author}{\bibfnamefont{L.}~\bibnamefont{Adamczyk}} \bibnamefont{et~al.}
  (\bibinfo{collaboration}{STAR}), \bibinfo{journal}{Phys. Rev. Lett.}
  \textbf{\bibinfo{volume}{116}}, \bibinfo{pages}{132301}
  (\bibinfo{year}{2016}), \eprint{1511.06003}.

\bibitem[{\citenamefont{Aghasyan et~al.}(2017)}]{Aghasyan:2017jop}
\bibinfo{author}{\bibfnamefont{M.}~\bibnamefont{Aghasyan}} \bibnamefont{et~al.}
  (\bibinfo{collaboration}{COMPASS}), \bibinfo{journal}{Phys. Rev. Lett.}
  \textbf{\bibinfo{volume}{119}}, \bibinfo{pages}{112002}
  (\bibinfo{year}{2017}), \eprint{1704.00488}.

\bibitem[{\citenamefont{Bomhof and Mulders}(2007)}]{Bomhof:2006ra}
\bibinfo{author}{\bibfnamefont{C.~J.} \bibnamefont{Bomhof}} \bibnamefont{and}
  \bibinfo{author}{\bibfnamefont{P.~J.} \bibnamefont{Mulders}},
  \bibinfo{journal}{JHEP} \textbf{\bibinfo{volume}{02}}, \bibinfo{pages}{029}
  (\bibinfo{year}{2007}), \eprint{hep-ph/0609206}.

\bibitem[{\citenamefont{Buffing et~al.}(2013)\citenamefont{Buffing, Mukherjee,
  and Mulders}}]{Buffing:2013kca}
\bibinfo{author}{\bibfnamefont{M.~G.~A.} \bibnamefont{Buffing}},
  \bibinfo{author}{\bibfnamefont{A.}~\bibnamefont{Mukherjee}},
  \bibnamefont{and} \bibinfo{author}{\bibfnamefont{P.~J.}
  \bibnamefont{Mulders}}, \bibinfo{journal}{Phys. Rev.}
  \textbf{\bibinfo{volume}{D88}}, \bibinfo{pages}{054027}
  (\bibinfo{year}{2013}), \eprint{1306.5897}.

\bibitem[{\citenamefont{Kishore and Mukherjee}(2019)}]{Kishore:2018ugo}
\bibinfo{author}{\bibfnamefont{R.}~\bibnamefont{Kishore}} \bibnamefont{and}
  \bibinfo{author}{\bibfnamefont{A.}~\bibnamefont{Mukherjee}},
  \bibinfo{journal}{Phys. Rev.} \textbf{\bibinfo{volume}{D99}},
  \bibinfo{pages}{054012} (\bibinfo{year}{2019}), \eprint{1811.07495}.

\bibitem[{\citenamefont{Lansberg et~al.}(2017)\citenamefont{Lansberg, Pisano,
  and Schlegel}}]{Lansberg:2017tlc}
\bibinfo{author}{\bibfnamefont{J.-P.} \bibnamefont{Lansberg}},
  \bibinfo{author}{\bibfnamefont{C.}~\bibnamefont{Pisano}}, \bibnamefont{and}
  \bibinfo{author}{\bibfnamefont{M.}~\bibnamefont{Schlegel}},
  \bibinfo{journal}{Nucl. Phys.} \textbf{\bibinfo{volume}{B920}},
  \bibinfo{pages}{192} (\bibinfo{year}{2017}), \eprint{1702.00305}.

\bibitem[{\citenamefont{Rajesh et~al.}(2018)\citenamefont{Rajesh, Kishore, and
  Mukherjee}}]{Rajesh:2018qks}
\bibinfo{author}{\bibfnamefont{S.}~\bibnamefont{Rajesh}},
  \bibinfo{author}{\bibfnamefont{R.}~\bibnamefont{Kishore}}, \bibnamefont{and}
  \bibinfo{author}{\bibfnamefont{A.}~\bibnamefont{Mukherjee}},
  \bibinfo{journal}{Phys. Rev.} \textbf{\bibinfo{volume}{D98}},
  \bibinfo{pages}{014007} (\bibinfo{year}{2018}), \eprint{1802.10359}.

\bibitem[{\citenamefont{Mukherjee and
  Rajesh}(2017{\natexlab{a}})}]{Mukherjee:2016qxa}
\bibinfo{author}{\bibfnamefont{A.}~\bibnamefont{Mukherjee}} \bibnamefont{and}
  \bibinfo{author}{\bibfnamefont{S.}~\bibnamefont{Rajesh}},
  \bibinfo{journal}{Eur. Phys. J.} \textbf{\bibinfo{volume}{C77}},
  \bibinfo{pages}{854} (\bibinfo{year}{2017}{\natexlab{a}}),
  \eprint{1609.05596}.

\bibitem[{\citenamefont{D'Alesio
  et~al.}(2017{\natexlab{a}})\citenamefont{D'Alesio, Murgia, Pisano, and
  Taels}}]{DAlesio:2017rzj}
\bibinfo{author}{\bibfnamefont{U.}~\bibnamefont{D'Alesio}},
  \bibinfo{author}{\bibfnamefont{F.}~\bibnamefont{Murgia}},
  \bibinfo{author}{\bibfnamefont{C.}~\bibnamefont{Pisano}}, \bibnamefont{and}
  \bibinfo{author}{\bibfnamefont{P.}~\bibnamefont{Taels}},
  \bibinfo{journal}{Phys. Rev.} \textbf{\bibinfo{volume}{D96}},
  \bibinfo{pages}{036011} (\bibinfo{year}{2017}{\natexlab{a}}),
  \eprint{1705.04169}.

\bibitem[{\citenamefont{Mukherjee and Rajesh}(2016)}]{Mukherjee:2015smo}
\bibinfo{author}{\bibfnamefont{A.}~\bibnamefont{Mukherjee}} \bibnamefont{and}
  \bibinfo{author}{\bibfnamefont{S.}~\bibnamefont{Rajesh}},
  \bibinfo{journal}{Phys. Rev.} \textbf{\bibinfo{volume}{D93}},
  \bibinfo{pages}{054018} (\bibinfo{year}{2016}), \eprint{1511.04319}.

\bibitem[{\citenamefont{Mukherjee and
  Rajesh}(2017{\natexlab{b}})}]{Mukherjee:2016cjw}
\bibinfo{author}{\bibfnamefont{A.}~\bibnamefont{Mukherjee}} \bibnamefont{and}
  \bibinfo{author}{\bibfnamefont{S.}~\bibnamefont{Rajesh}},
  \bibinfo{journal}{Phys. Rev.} \textbf{\bibinfo{volume}{D95}},
  \bibinfo{pages}{034039} (\bibinfo{year}{2017}{\natexlab{b}}),
  \eprint{1611.05974}.

\bibitem[{\citenamefont{Matoušek}(2016)}]{Matousek:2016xbl}
\bibinfo{author}{\bibfnamefont{J.}~\bibnamefont{Matoušek}}
  (\bibinfo{collaboration}{COMPASS}), \bibinfo{journal}{J. Phys. Conf. Ser.}
  \textbf{\bibinfo{volume}{678}}, \bibinfo{pages}{012050}
  (\bibinfo{year}{2016}).

\bibitem[{\citenamefont{D'Alesio
  et~al.}(2019{\natexlab{b}})\citenamefont{D'Alesio, Murgia, Pisano, and
  Taels}}]{DAlesio:2019qpk}
\bibinfo{author}{\bibfnamefont{U.}~\bibnamefont{D'Alesio}},
  \bibinfo{author}{\bibfnamefont{F.}~\bibnamefont{Murgia}},
  \bibinfo{author}{\bibfnamefont{C.}~\bibnamefont{Pisano}}, \bibnamefont{and}
  \bibinfo{author}{\bibfnamefont{P.}~\bibnamefont{Taels}}
  (\bibinfo{year}{2019}{\natexlab{b}}), \eprint{1908.00446}.

\bibitem[{\citenamefont{D'Alesio
  et~al.}(2017{\natexlab{b}})\citenamefont{D'Alesio, Flore, and
  Murgia}}]{DAlesio:2017nrd}
\bibinfo{author}{\bibfnamefont{U.}~\bibnamefont{D'Alesio}},
  \bibinfo{author}{\bibfnamefont{C.}~\bibnamefont{Flore}}, \bibnamefont{and}
  \bibinfo{author}{\bibfnamefont{F.}~\bibnamefont{Murgia}},
  \bibinfo{journal}{Phys. Rev.} \textbf{\bibinfo{volume}{D95}},
  \bibinfo{pages}{094002} (\bibinfo{year}{2017}{\natexlab{b}}),
  \eprint{1701.01148}.

\bibitem[{\citenamefont{Godbole et~al.}(2012)\citenamefont{Godbole, Misra,
  Mukherjee, and Rawoot}}]{Godbole:2012bx}
\bibinfo{author}{\bibfnamefont{R.~M.} \bibnamefont{Godbole}},
  \bibinfo{author}{\bibfnamefont{A.}~\bibnamefont{Misra}},
  \bibinfo{author}{\bibfnamefont{A.}~\bibnamefont{Mukherjee}},
  \bibnamefont{and} \bibinfo{author}{\bibfnamefont{V.~S.}
  \bibnamefont{Rawoot}}, \bibinfo{journal}{Phys. Rev.}
  \textbf{\bibinfo{volume}{D85}}, \bibinfo{pages}{094013}
  (\bibinfo{year}{2012}), \eprint{1201.1066}.

\bibitem[{\citenamefont{Godbole et~al.}(2013)\citenamefont{Godbole, Misra,
  Mukherjee, and Rawoot}}]{Godbole:2013bca}
\bibinfo{author}{\bibfnamefont{R.~M.} \bibnamefont{Godbole}},
  \bibinfo{author}{\bibfnamefont{A.}~\bibnamefont{Misra}},
  \bibinfo{author}{\bibfnamefont{A.}~\bibnamefont{Mukherjee}},
  \bibnamefont{and} \bibinfo{author}{\bibfnamefont{V.~S.}
  \bibnamefont{Rawoot}}, \bibinfo{journal}{Phys. Rev.}
  \textbf{\bibinfo{volume}{D88}}, \bibinfo{pages}{014029}
  (\bibinfo{year}{2013}), \eprint{1304.2584}.

\bibitem[{\citenamefont{Echevarria}(2019)}]{Echevarria:2019ynx}
\bibinfo{author}{\bibfnamefont{M.~G.} \bibnamefont{Echevarria}}
  (\bibinfo{year}{2019}), \eprint{1907.06494}.

\bibitem[{\citenamefont{Carlson and Suaya}(1976)}]{Carlson:1976cd}
\bibinfo{author}{\bibfnamefont{C.~E.} \bibnamefont{Carlson}} \bibnamefont{and}
  \bibinfo{author}{\bibfnamefont{R.}~\bibnamefont{Suaya}},
  \bibinfo{journal}{Phys. Rev.} \textbf{\bibinfo{volume}{D14}},
  \bibinfo{pages}{3115} (\bibinfo{year}{1976}).

\bibitem[{\citenamefont{Berger and Jones}(1981)}]{Berger:1980ni}
\bibinfo{author}{\bibfnamefont{E.~L.} \bibnamefont{Berger}} \bibnamefont{and}
  \bibinfo{author}{\bibfnamefont{D.~L.} \bibnamefont{Jones}},
  \bibinfo{journal}{Phys. Rev.} \textbf{\bibinfo{volume}{D23}},
  \bibinfo{pages}{1521} (\bibinfo{year}{1981}).

\bibitem[{\citenamefont{Baier and Ruckl}(1981)}]{Baier:1981uk}
\bibinfo{author}{\bibfnamefont{R.}~\bibnamefont{Baier}} \bibnamefont{and}
  \bibinfo{author}{\bibfnamefont{R.}~\bibnamefont{Ruckl}},
  \bibinfo{journal}{Phys. Lett.} \textbf{\bibinfo{volume}{102B}},
  \bibinfo{pages}{364} (\bibinfo{year}{1981}).

\bibitem[{\citenamefont{Baier and Ruckl}(1982)}]{Baier:1981zz}
\bibinfo{author}{\bibfnamefont{R.}~\bibnamefont{Baier}} \bibnamefont{and}
  \bibinfo{author}{\bibfnamefont{R.}~\bibnamefont{Ruckl}},
  \bibinfo{journal}{Nucl. Phys.} \textbf{\bibinfo{volume}{B201}},
  \bibinfo{pages}{1} (\bibinfo{year}{1982}).

\bibitem[{\citenamefont{Braaten and Fleming}(1995)}]{Braaten:1994vv}
\bibinfo{author}{\bibfnamefont{E.}~\bibnamefont{Braaten}} \bibnamefont{and}
  \bibinfo{author}{\bibfnamefont{S.}~\bibnamefont{Fleming}},
  \bibinfo{journal}{Phys. Rev. Lett.} \textbf{\bibinfo{volume}{74}},
  \bibinfo{pages}{3327} (\bibinfo{year}{1995}), \eprint{hep-ph/9411365}.

\bibitem[{\citenamefont{Cho and Leibovich}(1996{\natexlab{a}})}]{Cho:1995vh}
\bibinfo{author}{\bibfnamefont{P.~L.} \bibnamefont{Cho}} \bibnamefont{and}
  \bibinfo{author}{\bibfnamefont{A.~K.} \bibnamefont{Leibovich}},
  \bibinfo{journal}{Phys. Rev.} \textbf{\bibinfo{volume}{D53}},
  \bibinfo{pages}{150} (\bibinfo{year}{1996}{\natexlab{a}}),
  \eprint{hep-ph/9505329}.

\bibitem[{\citenamefont{Cho and Leibovich}(1996{\natexlab{b}})}]{Cho:1995ce}
\bibinfo{author}{\bibfnamefont{P.~L.} \bibnamefont{Cho}} \bibnamefont{and}
  \bibinfo{author}{\bibfnamefont{A.~K.} \bibnamefont{Leibovich}},
  \bibinfo{journal}{Phys. Rev.} \textbf{\bibinfo{volume}{D53}},
  \bibinfo{pages}{6203} (\bibinfo{year}{1996}{\natexlab{b}}),
  \eprint{hep-ph/9511315}.

\bibitem[{\citenamefont{Lepage et~al.}(1992)\citenamefont{Lepage, Magnea,
  Nakhleh, Magnea, and Hornbostel}}]{Lepage:1992tx}
\bibinfo{author}{\bibfnamefont{G.~P.} \bibnamefont{Lepage}},
  \bibinfo{author}{\bibfnamefont{L.}~\bibnamefont{Magnea}},
  \bibinfo{author}{\bibfnamefont{C.}~\bibnamefont{Nakhleh}},
  \bibinfo{author}{\bibfnamefont{U.}~\bibnamefont{Magnea}}, \bibnamefont{and}
  \bibinfo{author}{\bibfnamefont{K.}~\bibnamefont{Hornbostel}},
  \bibinfo{journal}{Phys. Rev.} \textbf{\bibinfo{volume}{D46}},
  \bibinfo{pages}{4052} (\bibinfo{year}{1992}), \eprint{hep-lat/9205007}.

\bibitem[{\citenamefont{Abe et~al.}(1997)}]{Abe:1997jz}
\bibinfo{author}{\bibfnamefont{F.}~\bibnamefont{Abe}} \bibnamefont{et~al.}
  (\bibinfo{collaboration}{CDF}), \bibinfo{journal}{Phys. Rev. Lett.}
  \textbf{\bibinfo{volume}{79}}, \bibinfo{pages}{572} (\bibinfo{year}{1997}).

\bibitem[{\citenamefont{Acosta et~al.}(2005)}]{Acosta:2004yw}
\bibinfo{author}{\bibfnamefont{D.}~\bibnamefont{Acosta}} \bibnamefont{et~al.}
  (\bibinfo{collaboration}{CDF}), \bibinfo{journal}{Phys. Rev.}
  \textbf{\bibinfo{volume}{D71}}, \bibinfo{pages}{032001}
  (\bibinfo{year}{2005}), \eprint{hep-ex/0412071}.

\bibitem[{\citenamefont{Adloff et~al.}(2002)}]{Adloff:2002ex}
\bibinfo{author}{\bibfnamefont{C.}~\bibnamefont{Adloff}} \bibnamefont{et~al.}
  (\bibinfo{collaboration}{H1}), \bibinfo{journal}{Eur. Phys. J.}
  \textbf{\bibinfo{volume}{C25}}, \bibinfo{pages}{25} (\bibinfo{year}{2002}),
  \eprint{hep-ex/0205064}.

\bibitem[{\citenamefont{Aaron et~al.}(2010)}]{Aaron:2010gz}
\bibinfo{author}{\bibfnamefont{F.~D.} \bibnamefont{Aaron}} \bibnamefont{et~al.}
  (\bibinfo{collaboration}{H1}), \bibinfo{journal}{Eur. Phys. J.}
  \textbf{\bibinfo{volume}{C68}}, \bibinfo{pages}{401} (\bibinfo{year}{2010}),
  \eprint{1002.0234}.

\bibitem[{\citenamefont{Chekanov et~al.}(2003)}]{Chekanov:2002at}
\bibinfo{author}{\bibfnamefont{S.}~\bibnamefont{Chekanov}} \bibnamefont{et~al.}
  (\bibinfo{collaboration}{ZEUS}), \bibinfo{journal}{Eur. Phys. J.}
  \textbf{\bibinfo{volume}{C27}}, \bibinfo{pages}{173} (\bibinfo{year}{2003}),
  \eprint{hep-ex/0211011}.

\bibitem[{\citenamefont{Abramowicz et~al.}(2013)}]{Abramowicz:2012dh}
\bibinfo{author}{\bibfnamefont{H.}~\bibnamefont{Abramowicz}}
  \bibnamefont{et~al.} (\bibinfo{collaboration}{ZEUS}), \bibinfo{journal}{JHEP}
  \textbf{\bibinfo{volume}{02}}, \bibinfo{pages}{071} (\bibinfo{year}{2013}),
  \eprint{1211.6946}.

\bibitem[{\citenamefont{Frixione et~al.}(1993)\citenamefont{Frixione, Mangano,
  Nason, and Ridolfi}}]{Frixione:1993yw}
\bibinfo{author}{\bibfnamefont{S.}~\bibnamefont{Frixione}},
  \bibinfo{author}{\bibfnamefont{M.~L.} \bibnamefont{Mangano}},
  \bibinfo{author}{\bibfnamefont{P.}~\bibnamefont{Nason}}, \bibnamefont{and}
  \bibinfo{author}{\bibfnamefont{G.}~\bibnamefont{Ridolfi}},
  \bibinfo{journal}{Phys. Lett.} \textbf{\bibinfo{volume}{B319}},
  \bibinfo{pages}{339} (\bibinfo{year}{1993}), \eprint{hep-ph/9310350}.

\bibitem[{\citenamefont{Boer et~al.}(2016)\citenamefont{Boer, Mulders, Pisano,
  and Zhou}}]{Boer:2016fqd}
\bibinfo{author}{\bibfnamefont{D.}~\bibnamefont{Boer}},
  \bibinfo{author}{\bibfnamefont{P.~J.} \bibnamefont{Mulders}},
  \bibinfo{author}{\bibfnamefont{C.}~\bibnamefont{Pisano}}, \bibnamefont{and}
  \bibinfo{author}{\bibfnamefont{J.}~\bibnamefont{Zhou}},
  \bibinfo{journal}{JHEP} \textbf{\bibinfo{volume}{08}}, \bibinfo{pages}{001}
  (\bibinfo{year}{2016}), \eprint{1605.07934}.

\bibitem[{\citenamefont{Bacchetta et~al.}(2004)\citenamefont{Bacchetta,
  D'Alesio, Diehl, and Miller}}]{Bacchetta:2004jz}
\bibinfo{author}{\bibfnamefont{A.}~\bibnamefont{Bacchetta}},
  \bibinfo{author}{\bibfnamefont{U.}~\bibnamefont{D'Alesio}},
  \bibinfo{author}{\bibfnamefont{M.}~\bibnamefont{Diehl}}, \bibnamefont{and}
  \bibinfo{author}{\bibfnamefont{C.~A.} \bibnamefont{Miller}},
  \bibinfo{journal}{Phys. Rev.} \textbf{\bibinfo{volume}{D70}},
  \bibinfo{pages}{117504} (\bibinfo{year}{2004}), \eprint{hep-ph/0410050}.

\bibitem[{\citenamefont{Bacchetta et~al.}(2007)\citenamefont{Bacchetta, Bomhof,
  D'Alesio, Mulders, and Murgia}}]{Bacchetta:2007sz}
\bibinfo{author}{\bibfnamefont{A.}~\bibnamefont{Bacchetta}},
  \bibinfo{author}{\bibfnamefont{C.}~\bibnamefont{Bomhof}},
  \bibinfo{author}{\bibfnamefont{U.}~\bibnamefont{D'Alesio}},
  \bibinfo{author}{\bibfnamefont{P.~J.} \bibnamefont{Mulders}},
  \bibnamefont{and} \bibinfo{author}{\bibfnamefont{F.}~\bibnamefont{Murgia}},
  \bibinfo{journal}{Phys. Rev. Lett.} \textbf{\bibinfo{volume}{99}},
  \bibinfo{pages}{212002} (\bibinfo{year}{2007}), \eprint{hep-ph/0703153}.

\bibitem[{\citenamefont{Collins}(2013)}]{jcollins}
\bibinfo{author}{\bibfnamefont{J.}~\bibnamefont{Collins}},
  \emph{\bibinfo{title}{{Foundations of perturbative QCD}}}
  (\bibinfo{publisher}{Cambridge University Press}, \bibinfo{year}{2013}).

\bibitem[{\citenamefont{Aybat et~al.}(2012{\natexlab{b}})\citenamefont{Aybat,
  Collins, Qiu, and Rogers}}]{aybat}
\bibinfo{author}{\bibfnamefont{S.~M.} \bibnamefont{Aybat}},
  \bibinfo{author}{\bibfnamefont{J.~C.} \bibnamefont{Collins}},
  \bibinfo{author}{\bibfnamefont{J.-W.} \bibnamefont{Qiu}}, \bibnamefont{and}
  \bibinfo{author}{\bibfnamefont{T.~C.} \bibnamefont{Rogers}},
  \bibinfo{journal}{Phys. Rev.} \textbf{\bibinfo{volume}{D85}},
  \bibinfo{pages}{034043} (\bibinfo{year}{2012}{\natexlab{b}}),
  \eprint{1110.6428}.

\bibitem[{\citenamefont{Ji et~al.}(2006)\citenamefont{Ji, Qiu, Vogelsang, and
  Yuan}}]{Ji:2006ub}
\bibinfo{author}{\bibfnamefont{X.}~\bibnamefont{Ji}},
  \bibinfo{author}{\bibfnamefont{J.-W.} \bibnamefont{Qiu}},
  \bibinfo{author}{\bibfnamefont{W.}~\bibnamefont{Vogelsang}},
  \bibnamefont{and} \bibinfo{author}{\bibfnamefont{F.}~\bibnamefont{Yuan}},
  \bibinfo{journal}{Phys. Rev. Lett.} \textbf{\bibinfo{volume}{97}},
  \bibinfo{pages}{082002} (\bibinfo{year}{2006}), \eprint{hep-ph/0602239}.

\bibitem[{\citenamefont{Kouvaris et~al.}(2006)\citenamefont{Kouvaris, Qiu,
  Vogelsang, and Yuan}}]{Kouvaris:2006zy}
\bibinfo{author}{\bibfnamefont{C.}~\bibnamefont{Kouvaris}},
  \bibinfo{author}{\bibfnamefont{J.-W.} \bibnamefont{Qiu}},
  \bibinfo{author}{\bibfnamefont{W.}~\bibnamefont{Vogelsang}},
  \bibnamefont{and} \bibinfo{author}{\bibfnamefont{F.}~\bibnamefont{Yuan}},
  \bibinfo{journal}{Phys. Rev.} \textbf{\bibinfo{volume}{D74}},
  \bibinfo{pages}{114013} (\bibinfo{year}{2006}), \eprint{hep-ph/0609238}.

\bibitem[{\citenamefont{Boer and Vogelsang}(2004)}]{Boer:2003tx}
\bibinfo{author}{\bibfnamefont{D.}~\bibnamefont{Boer}} \bibnamefont{and}
  \bibinfo{author}{\bibfnamefont{W.}~\bibnamefont{Vogelsang}},
  \bibinfo{journal}{Phys. Rev.} \textbf{\bibinfo{volume}{D69}},
  \bibinfo{pages}{094025} (\bibinfo{year}{2004}), \eprint{hep-ph/0312320}.

\bibitem[{\citenamefont{Yuan}(2008)}]{Yuan:2008vn}
\bibinfo{author}{\bibfnamefont{F.}~\bibnamefont{Yuan}}, \bibinfo{journal}{Phys.
  Rev.} \textbf{\bibinfo{volume}{D78}}, \bibinfo{pages}{014024}
  (\bibinfo{year}{2008}), \eprint{0801.4357}.

\bibitem[{\citenamefont{Chao et~al.}(2012)\citenamefont{Chao, Ma, Shao, Wang,
  and Zhang}}]{Chao:2012iv}
\bibinfo{author}{\bibfnamefont{K.-T.} \bibnamefont{Chao}},
  \bibinfo{author}{\bibfnamefont{Y.-Q.} \bibnamefont{Ma}},
  \bibinfo{author}{\bibfnamefont{H.-S.} \bibnamefont{Shao}},
  \bibinfo{author}{\bibfnamefont{K.}~\bibnamefont{Wang}}, \bibnamefont{and}
  \bibinfo{author}{\bibfnamefont{Y.-J.} \bibnamefont{Zhang}},
  \bibinfo{journal}{Phys. Rev. Lett.} \textbf{\bibinfo{volume}{108}},
  \bibinfo{pages}{242004} (\bibinfo{year}{2012}), \eprint{1201.2675}.

\bibitem[{\citenamefont{Li and Liu}(1998)}]{Li:1996jk}
\bibinfo{author}{\bibfnamefont{Y.-d.} \bibnamefont{Li}} \bibnamefont{and}
  \bibinfo{author}{\bibfnamefont{L.-s.} \bibnamefont{Liu}},
  \bibinfo{journal}{Commun. Theor. Phys.} \textbf{\bibinfo{volume}{29}},
  \bibinfo{pages}{99} (\bibinfo{year}{1998}).

\bibitem[{\citenamefont{Butenschoen and Kniehl}(2010)}]{Butenschoen:2009zy}
\bibinfo{author}{\bibfnamefont{M.}~\bibnamefont{Butenschoen}} \bibnamefont{and}
  \bibinfo{author}{\bibfnamefont{B.~A.} \bibnamefont{Kniehl}},
  \bibinfo{journal}{Phys. Rev. Lett.} \textbf{\bibinfo{volume}{104}},
  \bibinfo{pages}{072001} (\bibinfo{year}{2010}), \eprint{0909.2798}.

\bibitem[{\citenamefont{Artoisenet et~al.}(2009)\citenamefont{Artoisenet,
  Campbell, Maltoni, and Tramontano}}]{Artoisenet:2009xh}
\bibinfo{author}{\bibfnamefont{P.}~\bibnamefont{Artoisenet}},
  \bibinfo{author}{\bibfnamefont{J.~M.} \bibnamefont{Campbell}},
  \bibinfo{author}{\bibfnamefont{F.}~\bibnamefont{Maltoni}}, \bibnamefont{and}
  \bibinfo{author}{\bibfnamefont{F.}~\bibnamefont{Tramontano}},
  \bibinfo{journal}{Phys. Rev. Lett.} \textbf{\bibinfo{volume}{102}},
  \bibinfo{pages}{142001} (\bibinfo{year}{2009}), \eprint{0901.4352}.

\bibitem[{\citenamefont{Buckley et~al.}(2015)\citenamefont{Buckley, Ferrando,
  Lloyd, Nordström, Page, Rüfenacht, Schönherr, and Watt}}]{Buckley:2014ana}
\bibinfo{author}{\bibfnamefont{A.}~\bibnamefont{Buckley}},
  \bibinfo{author}{\bibfnamefont{J.}~\bibnamefont{Ferrando}},
  \bibinfo{author}{\bibfnamefont{S.}~\bibnamefont{Lloyd}},
  \bibinfo{author}{\bibfnamefont{K.}~\bibnamefont{Nordström}},
  \bibinfo{author}{\bibfnamefont{B.}~\bibnamefont{Page}},
  \bibinfo{author}{\bibfnamefont{M.}~\bibnamefont{Rüfenacht}},
  \bibinfo{author}{\bibfnamefont{M.}~\bibnamefont{Schönherr}},
  \bibnamefont{and} \bibinfo{author}{\bibfnamefont{G.}~\bibnamefont{Watt}},
  \bibinfo{journal}{Eur. Phys. J.} \textbf{\bibinfo{volume}{C75}},
  \bibinfo{pages}{132} (\bibinfo{year}{2015}), \eprint{1412.7420}.

\end{thebibliography}
\end{document}